\documentclass[trackchanges]{aastex7}
\usepackage{siunitx} 
\usepackage{graphicx}   
\usepackage{subcaption}  
\usepackage{hyperref}    
\usepackage{amsmath}
\usepackage{booktabs}  
\usepackage{xcolor}

\begin{document}


\title{Identifying Quasi-Periodic Micropulses in Pulsars with FAST Using Convolutional 
Neural Networks}

\author[0009-0005-1325-6421]{S.D. Wang}
\altaffiliation{These authors contributed equally to this work.}
\affiliation{Guizhou Normal University, Guiyang 550001, People’s Republic of China}
\email{232100070279@gznu.edu.cn} 

\author{H. Liu$^\dagger$}
\affiliation{Guizhou Normal University, Guiyang 550001, People’s Republic of China}
\email{201907006@gznu.edu.cn} 

\author[0000-0002-1243-0476]{R.S. Zhao$^{\dagger\dagger}$}
\affiliation{Guizhou Normal University, Guiyang 550001, People’s Republic of China}
\email{201907007@gznu.edu.cn} 

\author{Baoqiang Lao}
\altaffiliation{These authors contributed equally to this work.}
\affiliation{Guangxi Key Laboratory of Brain-inspired Computing and Intelligent Chips, School of Electronic and Information Engineering, GuangXi Normal University, Guilin 541004, People's Republic of China}
\affiliation{Key Laboratory of Nonlinear Circuits and Optical Communications (GuangXi Normal University), Education Department of Guangxi Zhuang Autonomous Region, Guilin 541004, People's Republic of China}
\email{lbq@gxnu.edu.cn} 

\author{Y.K. Zhang}
\affiliation{National Astronomical Observatories, Chinese Academy of Sciences, 20A Datun Road, Chaoyang District, Beĳing 100101, People’s Republic of China}
\email{ykzhang@nao.cas.cn} 

\author{Y.F. Xiao}
\affiliation{ Sun Yat-sen University, Zhuhai, 519082, People’s Republic of China}
\email{xiaoyf53@mail2.sysu.edu.cn}

\author{P. Wang}
\affiliation{National Astronomical Observatories, Chinese Academy of Sciences, 20A Datun Road, Chaoyang District, Beĳing 100101, People’s Republic of China}
\affiliation{State Key Laboratory of Radio Astronomy and Technology, Chinese Academy of Sciences, A20 Datun Road, Chaoyang District, Beijing, 100101, People’s Republic of China}
\affiliation{Institute for Frontiers in Astronomy and Astrophysics, Beĳing Normal University, Beĳing 102206, People’s Republic of China}
\email{wangpei@nao.cas.cn} 

\author{D. Li}
\affiliation{National Astronomical Observatories, Chinese Academy of Sciences, 20A Datun Road, Chaoyang District, Beĳing 100101, People’s Republic of China}
\affiliation{Department of Astronomy, Tsinghua University, Beijing 100084, People’s Republic of China}
\affiliation{State Key Laboratory of Radio Astronomy and Technology, Chinese Academy of Sciences, A20 Datun Road, Chaoyang District, Beijing, 100101, P.R. People’s Republic of China}
\affiliation{Zhejiang Lab, Hangzhou, Zhejiang 311121, People’s Republic of China}
\email{dili@nao.cas.cn}

\author[0009-0007-3215-2964]{R.W. Tian}
\affiliation{Guizhou Normal University, Guiyang 550001, People’s Republic of China}
\email{}

\author{Z.F. Tu}
\affiliation{Guizhou Normal University, Guiyang 550001, People’s Republic of China}
\email{232100070278@gznu.edu.cn} 

\author{Q. Zhou}
\affiliation{Guizhou Normal University, Guiyang 550001, People’s Republic of China}
\email{242100070283@gznu.edu.cn} 

\author{Z.J. Zhang}
\affiliation{Guizhou Normal University, Guiyang 550001, People’s Republic of China}
\email{242100070281@gznu.edu.cn} 




\author{Q.J. Zhi}
\affiliation{Guizhou Normal University, Guiyang 550001, People’s Republic of China}
\email{} 

\author{S.J. Dang}
\affiliation{Guizhou Normal University, Guiyang 550001, People’s Republic of China}
\email{} 

\author{Yang Kun}
\affiliation{Minzu Normal University of Xingyi, Xingyi 562400, People’s Republic of China}
\email{yangkun@xynun.edu.cn}

\correspondingauthor{H. Liu}
\email{201907006@gznu.edu.cn}

\correspondingauthor{R.S. Zhao}
\email{201907007@gznu.edu.cn} 

\correspondingauthor{Baoqiang Lao}
\email{lbq@gxnu.edu.cn} 


\begin{abstract}


Quasi-periodic MicroPulses (QMP) are quasi-periodic microstructural features manifested in individual pulsar radio pulses, the study of which is crucial for understanding pulsar radiation mechanisms. Manual identification of QMP in large-scale pulsar single-pulse datasets remains highly inefficient. To address this, we propose a Dual-Stage Residual Network (DSR) that achieves automated QMP detection in FAST observational data through joint analysis of single-pulse profiles 
 and their Amplitude Distribution Profiles (ADP), defined as the power spectra of the autocorrelation function derivatives of the microstructure residuals. The model was trained on PSR B1933+16 data from 2019 (10,486 single pulses) and evaluated on manually annotated PSR B1933+16 data from 2020 (9,657 single pulses). DSR achieved 96.10\% recall and 95.85\% precision on the test set. This approach provides an automated pipeline for large-scale, reproducible QMP identification and establishes the foundation for in-depth investigation of their physical mechanisms.

\end{abstract}



\section{Introduction}


Pulsars are rapidly rotating and highly magnetized compact objects. During their rotation, when radiation beams emitted from the polar cap regions sweep across the line of sight of radio telescopes, pulse signals can be detected. Within a single rotation period, pulse intensity often exhibits rapid fluctuations known as microstructures \citep{craft1968Natur.218.1122C}, among which those showing quasi-periodicity are termed quasi-periodic micropulses (QMP) \citep{Hankins1971}. Following the initial discovery of QMP in PSR B0950+08 by \citep{Hankins1971}, subsequent studies have identified them in other pulsars \citep{Hankins1972,Cordes1990,Lange1998A&A...332..111L,Kramer2002MNRAS.334..523K,Kuzmin2003,Crossley2010ApJ...722.1908C,Mitra2015ApJ...806..236M,De2016ApJ...833L..10D,Wen2021ApJ...918...57W,Zhao2023,Tian2025ApJ...982..107T,LiWei2025}. 
With the advancement of observational techniques, the detection of QMP in pulsars has become increasingly precise. Taking pulsar J0826+2637 as an example, \citet{Lange1998A&A...332..111L} first revealed that its QMP lies within a broad range of 0.36--0.66 ms based on observations with the Effelsberg 100-m telescope; \citet{Mitra2015ApJ...806..236M} determined a median value of $0.72\pm0.45$ ms using the Arecibo telescope. Leveraging the exceptional sensitivity of China's Five-hundred-meter Aperture Spherical radio Telescope (FAST), \citet{LiWei2025} precisely measured the QMP to be $0.49^{+0.14}_{-0.10}$ ms. The notably reduced error range fully demonstrates the precision advantage of FAST in QMP detection.

Numerous studies \citep{Kramer2002MNRAS.334..523K, Mitra2015ApJ...806..236M, De2016ApJ...833L..10D, Liu2022MNRAS.513.4037L, Dang2024MNRAS.528.1213D} have demonstrated a significant correlation between the quasi-periodic microstructure period ($P_\mu$) and the pulsar's rotation period ($P$). \citet{Kramer2024NatAs...8..230K} further established through a systematic study of all types of radio-emitting neutron stars that $P_\mu$ and $P$ follow the linear relationship: $P_\mu = (0.94 \pm 0.04) \times P^{(0.97 \pm 0.05)}$. The core physical significance of investigating QMPs lies in uncovering the origin of their quasi-periodicity: on one hand, it may arise from electrodynamic fluctuations or plasma instabilities (e.g., tearing modes \citep{Cheng1979ApJ...229..348C, Kramer2024NatAs...8..230K}), neutron star vibrations \citep{Boriakoff1976ApJ...208L..43B, Clemens2004ApJ...609..340C}, or radiation transfer modulation effects \citep{Harding1979ApJ...233..317H}; on the other hand, it may relate to the geometric distribution of plasma beams in the magnetosphere, manifesting as radial emission spot distributions \citep{Kramer2002MNRAS.334..523K} or angular beam models \citep{De2016ApJ...833L..10D, Kramer2024NatAs...8..230K}. Understanding these mechanisms is crucial for elucidating plasma flows generated by non-steady spark discharges in pulsar polar cap regions \citep{Mitra2015ApJ...806..236M}, magnetospheric instabilities  \citep{Kramer2024NatAs...8..230K}, and the microdynamics of relativistic particle radiation\citep{Kramer2024NatAs...8..230K}. Consequently, there is a dual need: (1) to conduct in-depth statistical validation of the $P_\mu$-$P$ relationship using more pulsar samples exhibiting QMP, and (2) to utilize statistical analyses of the quasi-periodic characteristic value $P_\mu$ to provide critical data support for constructing theories of pulsar magnetospheric geometry.

Current QMP identification methods include autocorrelation function (ACF) analysis of raw single pulses \citep{Hankins1972, Cordes1979AuJPh..32....9C, De2016ApJ...833L..10D}, power spectrum analysis \citep{Lange1998A&A...332..111L, Popov2002}, and ACF analysis following microstructural signal separation \citep{Mitra2015ApJ...806..236M, Zhao2023, Tian2025ApJ...982..107T}. However, all methods require manual screening of individual pulses from massive datasets, introducing efficiency bottlenecks and human error. FAST's high sensitivity enables detection of more such signals, with its sky surveys producing $10^6$ to $10^7$ single pulses daily \citep{Nan2011IJMPD..20..989N}. This dramatic increase in data volume poses severe challenges for QMP detection. To address the dual challenges of data volume and signal rarity, developing robust automated detection pipelines has become imperative.

Machine learning has demonstrated unique advantages in astronomical big data processing \citep{Ball2010IJMPD..19.1049B, Baron2019arXiv190407248B, Ivezi2020sdmm.book.....I}. This capability has been directly applied to establish robust pipelines for pulsar candidate screening \citep{Wang2025AstTI...2...27W}. 
Evolution of methods for screening pulsar candidate signals began with early feature engineering, such as the 12-dimensional ANN by \citet{Eatough2010MNRAS.407.2443E} and the 22-dimensional ANN by \citet{Bates2012}. Subsequently, more streamlined approaches emerged, including the minimalist six-feature classifier of \citet{Morello2014MNRAS.443.1651M} and the real-time decision trees of \citet{Lyon2016MNRAS.459.1104L}. The deep learning era then brought major advances—such as the pulsar image classification system (PICS) by \citet{Zhu2014ApJ...781..117Z} (later optimized with ResNet architectures by \citet{Wang2019SCPMA..6259507W}), concatenated CNNs \citep{Zeng2020}, generative adversarial networks \citep{Yin2023ApJS..264....2Y}, and hybrid CNN-attention models \citep{Cai2023RAA....23j4005C}. While existing methods predominantly distinguish pulsars from radio frequency interference, \citet{Amarnath2025asi..confO..33A} pioneered QMP-specific detection (without providing quantitative metrics). Notably, QMP identification shares core methodology with pulsar candidate recognition—both being binary classification tasks in image space—making machine learning equally viable for this application.

To address these needs, we propose a Dual-Stage Residual Network (DSR) that achieves automated QMP detection in FAST observational data through joint analysis of single-pulse profiles and ADP. The paper is structured as follows: Section \ref{sec:data} details data processing, feature analysis, and image preprocessing; Section \ref{sec:dsr-model} introduces the DSR architecture, experimental design and training results; Section \ref{sec:re} presents testing results; Section \ref{sec:dis} discusses experimental findings; and Section \ref{sec:con} provides concluding remarks.

\section{Data}\label{sec:data}

\subsection{Observational Data Processing}
This study analyzes archival FAST observational data covering three pulsars: PSR B1933+16 (2019-2022), PSR J0034-0721 (2021), and PSR J0304+1932 (2023). All observations utilized the central beam of the 19-beam receiver in tracking mode at 1250 MHz center frequency with 500 MHz bandwidth. Data were recorded in 8-bit PSRFITS search mode format with \qty{49.512}{\micro\second} sampling time.
Regarding the observational setup and data reduction, the observations used 4096 frequency channels for PSR B1933+16 and PSR J0304+1932, and 1024 channels for PSR J0034-0721, yielding channel resolutions of 0.122 MHz and 0.488 MHz, respectively. The data were processed using incoherent dedispersion. To evaluate whether the channel resolution was sufficient to neglect intra-channel dispersive smearing, we calculated the smearing timescale $\Delta t$ using the dispersion delay formula \citep{Lorimer2004hpa..book.....L}: 
\begin{equation}
\Delta t = 4.148808 \,\text{ms} \times \left[ \left( \frac{f_{\text{lo}}}{\text{GHz}} \right)^{-2} - \left( \frac{f_{\text{hi}}}{\text{GHz}} \right)^{-2} \right] \times \left( \frac{\text{DM}}{\text{cm}^{-3}  \text{pc}} \right)
\end{equation}
The results (Table~\ref{tab:pulsar_summary}) show that for all three pulsars, the maximum intra-channel smearing values are substantially smaller than the corresponding quasi-period scales of the micropulses, confirming that dispersive smearing does not affect the analysis of the micropulse structure.

The processing pipeline comprised three critical steps: First, single-pulse stacks were generated from dedispersed data using DSPSR software \citep{van2010asclsoft10006V}; Second, ephemerides were acquired via PSRCAT \citep{Manchester2005AJ....129.1993M}; Third, radio frequency interference (RFI) was excised in the frequency domain from the single-pulse stacks through preliminary cleaning of contaminated channels using the paz tool within the PSRCHIVE software package, followed by visual inspection and manual removal of residual interference signals \citep{Hotan2004PASA...21..302H}. In the final stage of data reduction, for the purpose of pulse profile analysis, we averaged the full frequency band into a single channel using the \texttt{pam} tool from the PSRCHIVE software package.
This yielded 46,131 clean single pulses. The 2019 data of PSR B1933+16 (10,486 pulses) were manually labeled and allocated for training and validation, while the 2020 data (9,657 pulses) served as the labeled test set. The remaining data formed the unlabeled test sets, including the 2021 (9,704 pulses) and 2022 (9,752 pulses) data of PSR B1933+16, the data of PSR J0034-0721 (3,118 pulses), and both the Dominant Component (DC) and Leading Component (LC) data of PSR J0304+1932 (3,414 pulses each). In summary, the manually annotated 2019 and 2020 data for PSR B1933+16 were used for model training, validation, and testing. The model's generalization capability was then evaluated on the remaining unlabeled data, where its QMP predictions were manually checked to determine the accuracy on these broader datasets. (Section~\ref{sec:test}). The observational information and corresponding pulse window longitude ranges are summarized in Table~\ref{tab:obs_data}, forming the basis for subsequent QMP classification.

\begin{deluxetable*}{ccccccc}[!ht]
\digitalasset
\tablewidth{0pt}
\tablecaption{Summary of pulsar parameters and dispersion delays \label{tab:pulsar_summary}}
\tablehead{
\colhead{PSR} & \colhead{DM} & \colhead{Channels} & \colhead{Channel} &\colhead{Low-freq} & \colhead{High-freq} &\colhead{Average} \\
\colhead{} & \colhead{(pc ${\rm cm}^{-3}$)} & \colhead{} & \colhead{resolution} &\colhead{$\Delta t$} & \colhead{$\Delta t$} &\colhead{QP} \\
\colhead{} & \colhead{} & \colhead{} & \colhead{(MHz)} &\colhead{($\mu$s)} & \colhead{($\mu$s)} &\colhead{$P_{\mu}$} \\
\colhead{} & \colhead{} & \colhead{} & \colhead{} &\colhead{} & \colhead{} &\colhead{($\mu$s)}
}
\startdata
B1933+16 & 158.52 & 4096 & 0.122 & 160.57 & 43.96 & 310.37 \\
J0304+1932 & 15.66 & 4096 & 0.122 & 15.86 & 4.34 & 1148.36 \\
J0034-0721 & 10.92 & 1024 & 0.488 & 44.25 & 11.37 & 711.22 \\
\enddata
\tablecomments{The frequency range for the three pulsars is from 1000 MHz to 1500 MHz with a total bandwidth of 500 MHz. The channel resolution is calculated as $\text{Channel Resolution} = \text{Total Bandwidth} / \text{Channels}$. The low-frequency range for $\Delta t$ calculation is: $f_{\text{lo}} = 1000$ MHz, $f_{\text{hi}} = 1000 + \text{Channel Resolution}$ MHz. The high-frequency range is: $f_{\text{lo}} = 1500 - \text{Channel Resolution}$ MHz, $f_{\text{hi}} = 1500$ MHz. The average quasi-period values ($P_\mu$) are computed from the microstructure quasi-periods measured in Table~\ref{tab:performance_Gaussian_Fitting}.}
\end{deluxetable*}

\begin{deluxetable*}{cccccc}[!ht]
\digitalasset
\tablewidth{0pt}
\tablecaption{Detailed observational information and pulse window longitude ranges for the three pulsars \label{tab:obs_data}}
\tablehead{
\colhead{PSR} & \colhead{Processed} & \colhead{Longitude} & \colhead{Bins} &\colhead{Bin Width} & \colhead{Observation} \\
\colhead{} & \colhead{pulses} & \colhead{range ($^\circ$)} & \colhead{} &\colhead{($\mu$s)} & \colhead{date}
}
\startdata
B1933+16 & 10,486 & -3.3--3.3 & 4096 & 87.53 & 2019-09-19 \\
B1933+16 & 9,657  & -3.5--3.0 & ... & ... & 2020-01-09 \\
B1933+16 & 9,704  & -3.4--3.2 & ... & ... & 2021-04-12 \\
B1933+16 & 9,752  & -3.4--3.3 & ... & ... & 2022-05-03 \\
J0034-0721 & 3,118  & -6.6--7.7  & ... & 230.21 & 2021-12-23  \\
J0304+1932 (DC) & 3,414  & -3.5--5.2 &  16384 &  84.69 & 2023-11-07 \\
J0304+1932 (LC) & 3,414  & -14.5--3.5   & ... & ... & ... \\
\enddata
\tablecomments{PSR J0304+1932 exhibits two subpulse components: Dominant Component (DC) and Leading Component (LC). In the average pulse profile, the dominant component is the component with higher intensity, while the leading component is the component that appears earlier in the pulse phase. The ellipsis (...) indicates that the value is identical to the one directly above it in the same column.}
\end{deluxetable*}

\subsection{QMP Classification Methodology}
\label{sec:classification}
\begin{figure*}[ht!]
  \centering
  \captionsetup[subfigure]{labelformat=simple}
  \renewcommand{\thesubfigure}{(\roman{subfigure})}
  \begin{subfigure}[b]{0.48\textwidth}
    \centering
    \includegraphics[width=\textwidth]{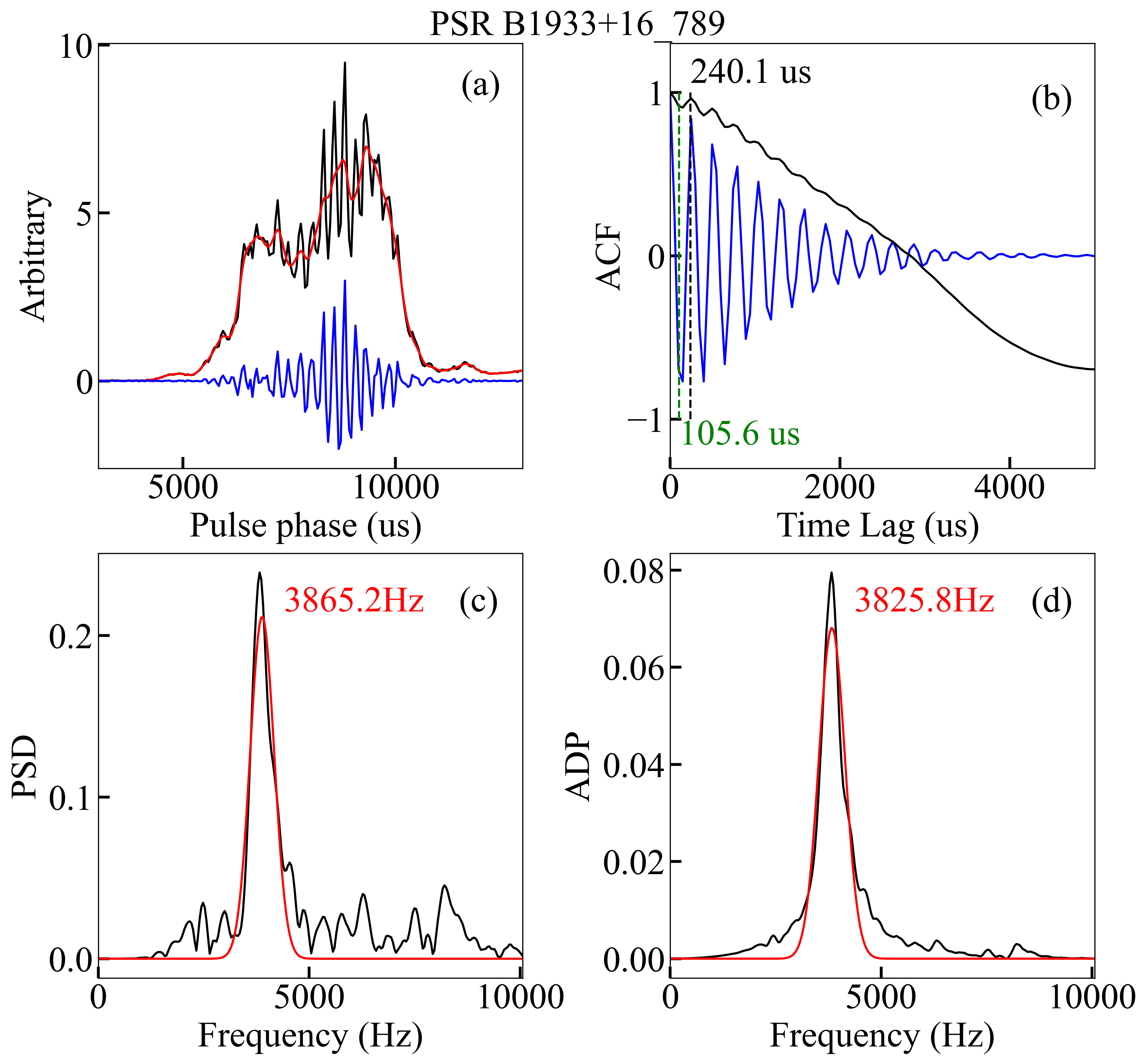}
    \caption{PSR B1933+16 pulse 789 (QMP)} 
    \label{fig:sub1}
  \end{subfigure}
  \hfill
  \begin{subfigure}[b]{0.48\textwidth}
    \centering
    \includegraphics[width=\textwidth]{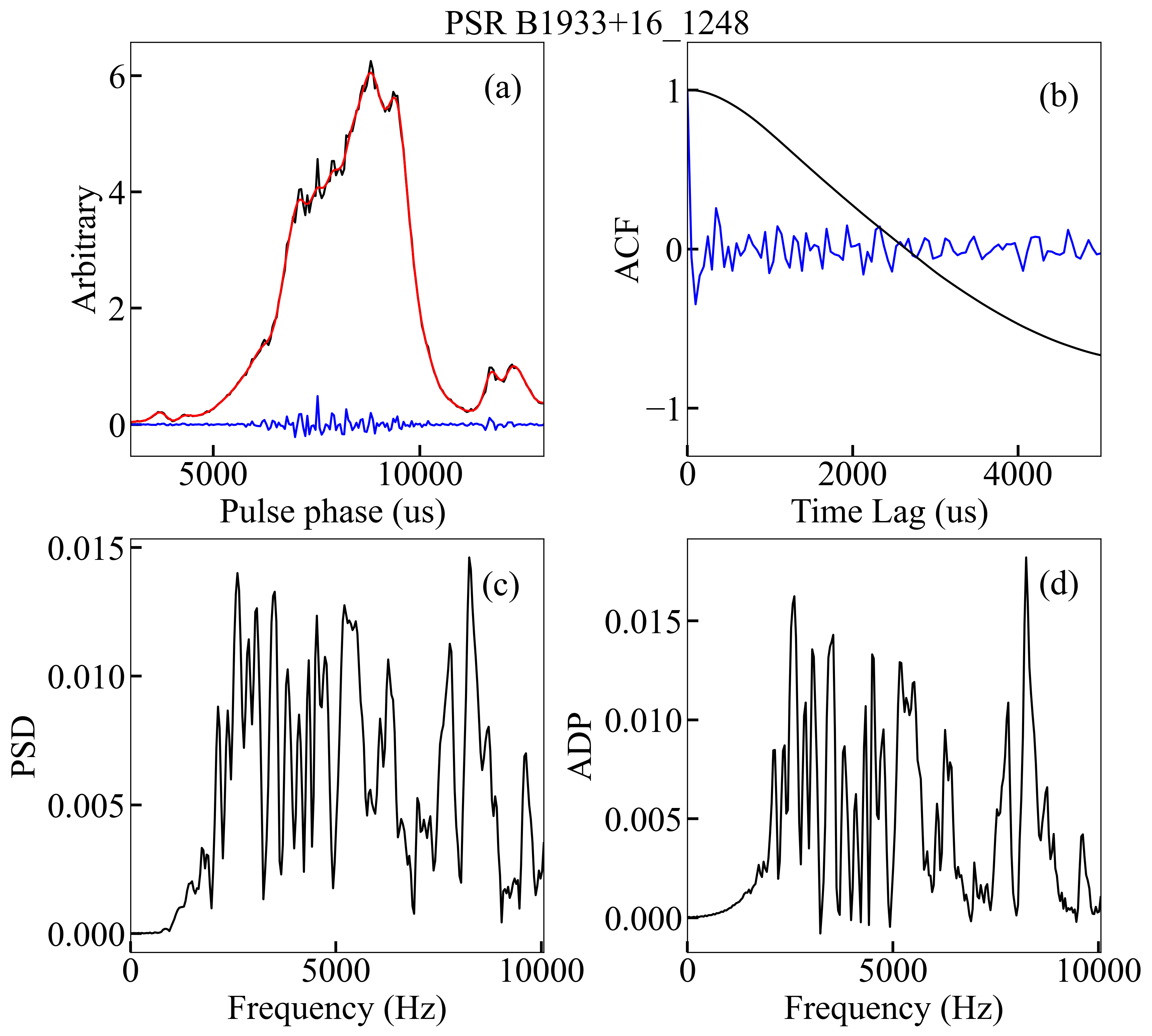}
    \caption{PSR B1933+16 pulse 1248 (NQMP)} 
    \label{fig:sub2}
  \end{subfigure}

\vspace{0.5cm}

\begin{subfigure}[b]{0.48\textwidth}
    \centering
    \includegraphics[width=\textwidth]{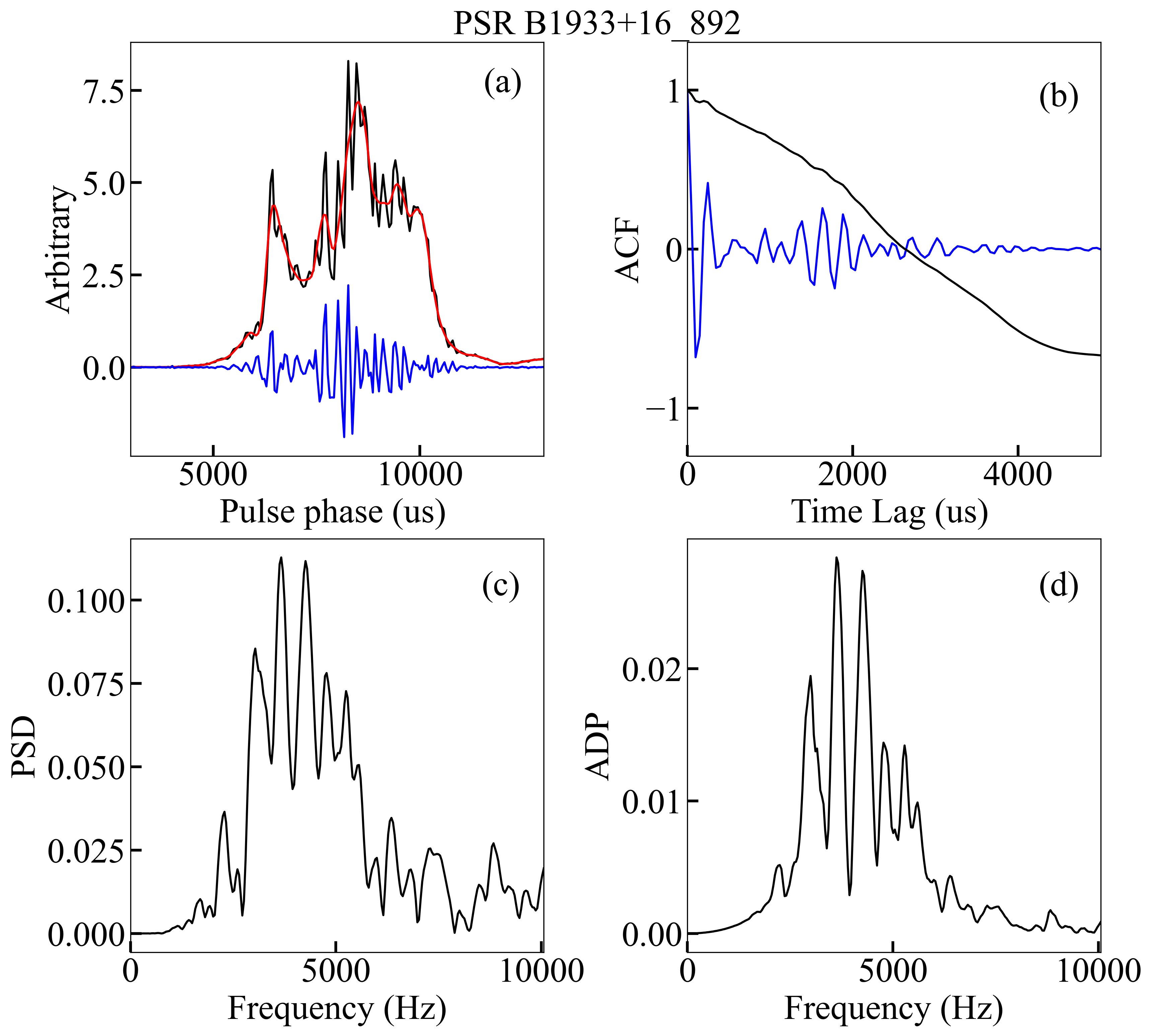}
    \caption{PSR B1933+16 pulse 892 (NQMP)} 
    \label{fig:sub3}
\end{subfigure}
\hfill
\begin{subfigure}[b]{0.48\textwidth}
    \centering
    \includegraphics[width=\textwidth]{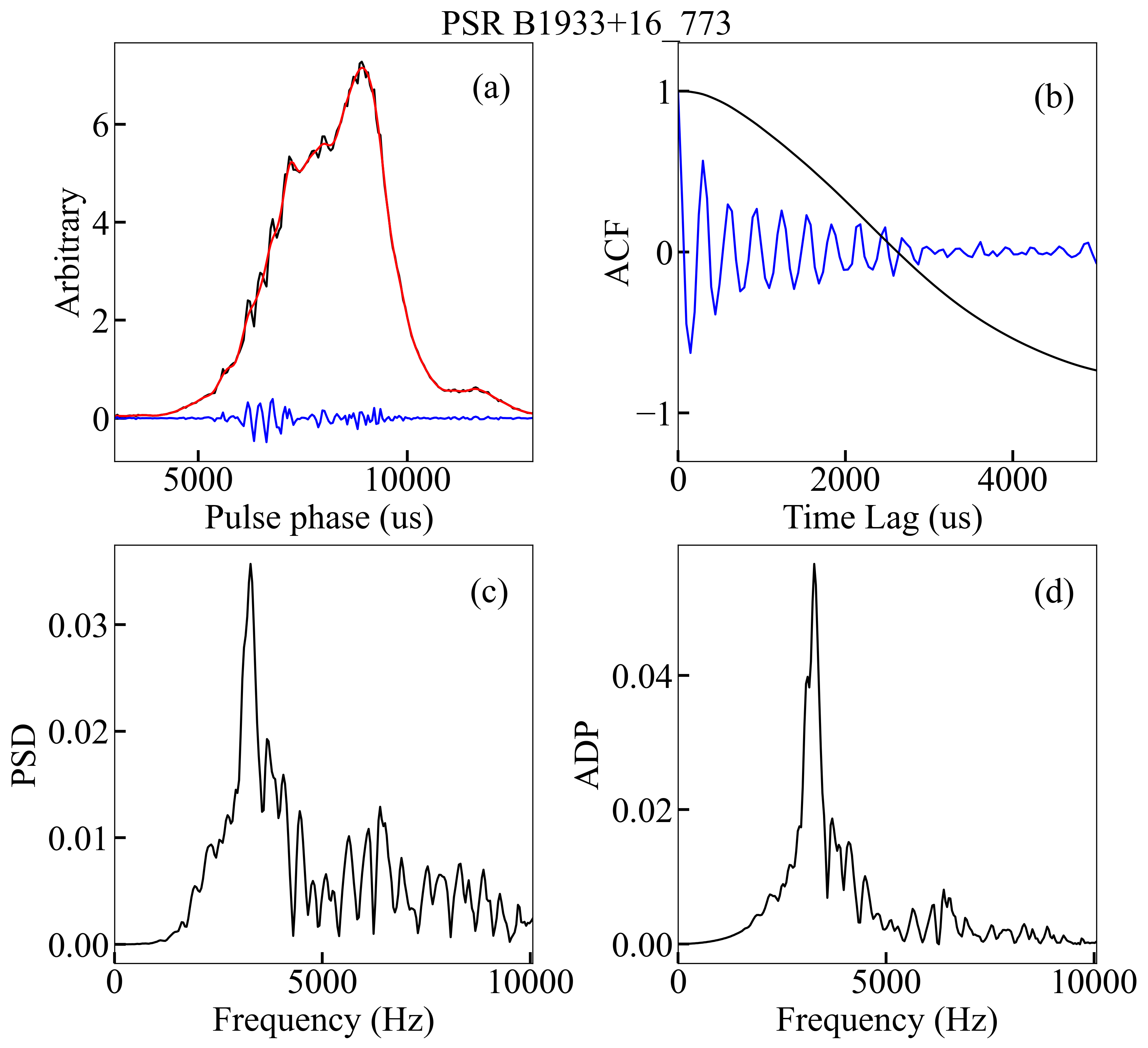}
    \caption{PSR B1933+16 pulse 773 (NQMP)} 
    \label{fig:sub4}
\end{subfigure}

\caption{Diagnostic plot comparison of QMP and NQMP in PSR B1933+16. (i) Pulse 789 with QMP: (a) Original subpulse data (black trace) exhibits pronounced and clustered quasi-periodic microstructure; (b) ACF of microstructure features (blue trace) displays significant periodicity; (c,d) Power spectra (PSD and ADP) show distinct unimodal signatures. (ii) Pulse 1248 with NQMP: (a) Original subpulse data (black trace) shows random fluctuations with indistinct microstructure; (b) ACF of microstructure features (blue trace) presents chaotic patterns; (c,d) Power spectra lack significant unimodal features. (iii) Pulse 892 with NQMP: (a) Original subpulse data (black trace) exhibits microstructure with clear oscillations; (b) Blue trace shows no periodicity; (c,d) Power spectra lack significant unimodal characteristics. (iv) Pulse 773 with NQMP: (a) Original subpulse data (black trace) lacks pronounced clustered microstructure, but contains intrinsic profile oscillations that were misidentified as QMP during microstructure extraction (blue line); thus (b) blue trace exhibits periodicity and (c,d) power spectra show significant unimodal features.}
  \label{fig:diagnostic} 
\end{figure*}

We employed the ACF and ADP method (\citet{Tian2025ApJ...982..107T}) to strictly classify single pulses into two categories (Figure~\ref{fig:diagnostic}): QMP and Non-quasi-periodic Micropulses (NQMP). We employed the ACF and ADP method (\citet{Tian2025ApJ...982..107T}) to strictly classify single pulses into two categories (Figure~\ref{fig:diagnostic}): QMP and Non-quasi-periodic Micropulses (NQMP). First, pulse window regions were extracted based on average pulse profiles, where their positions (expressed in longitude ranges) are given in Table~\ref{tab:obs_data}. Multicomponent pulses were processed separately because they are known to exhibit distinctive temporal characteristics, including different quasi-periodicities \citep{Soglasnov1981SvA....25..442S,Soglasnov1983SvA....27..169S,Popov2002}. 
Second, microstructure features were extracted within pulse windows (Figure~\ref{fig:sub1}a): the black trace shows original subpulse data, the red solid line represents the smoothed trend obtained through 5th-order polynomial fitting, with the goodness of fit assessed by requiring the extracted microstructure features to exhibit a peak-to-trough amplitude greater than 5 times the RMS noise level, and the blue solid line indicates extracted microstructure features. Third, power spectral density (PSD) of microstructure features and ADP were analyzed (black curves in Figure~\ref{fig:sub1}c,d) \citep{Lange1998A&A...332..111L}. 
Red curves show Gaussian fits, with PSD and ADP exhibiting significant peaks at 3865.2 Hz and 3825.8 Hz respectively, consistent within the 20\% tolerance threshold defined by \cite{Kramer2024NatAs...8..230K} with the \qty{240.1}{\si{\mu s}} quasi-period estimated from ACF (black dashed line in Figure~\ref{fig:sub1}b). Finally, the signal-to-noise ratio (SNR) of ADP was calculated as the ratio of the unimodal peak amplitude in the red Gaussian fit curve (Figure~\ref{fig:sub1}d) to the RMS of the power spectrum. This computational approach was validated on the 10,486-pulse dataset from PSR B1933+16 (2019): pulses with ADP SNR \textgreater 10 were classified as QMP candidates (784 pulses, 7.48\%), while the remaining 9,702 pulses (92.52\%) were categorized as NQMP.

Beyond the typical NQMPs shown in Figure~\ref{fig:sub2} (lacking distinct quasi-periodic microstructure in single-pulse profiles and showing multimodal ADP with SNR \textless 10), two special cases require differentiation: pulses exhibiting clustered quasi-periodic microstructure in time domain but multimodal ADP structure (Figure~\ref{fig:sub3}), which may indicate a mixture of multiple quasi-periodicities rather than absence of quasi-periodicity, and since we only select cases with single quasi-periodicity as QMPs, this confirms the insufficiency of time-domain analysis alone; and pulses showing unimodal ADP (SNR \textgreater 10) but insignificant clustered microstructure in time domain (Figure~\ref{fig:sub4}), where fluctuations may represent a subclass of microstructure. To ensure rigorous QMP selection, we retained only pulses simultaneously satisfying both criteria of pronounced clustered microstructure in time-domain (e.g., Figure~\ref{fig:sub1}a) and distinct unimodal ADP with SNR \textgreater 10. Table~\ref{tab:class_stats} shows quantitative distributions across four diagnostic categories for PSR B1933+16 (2019): 400 QMP pulses (3.82\%) and 10,086 NQMP pulses. This method identified 794 QMPs (8.22\%) in the 2020 dataset (9,657 pulses).

\begin{deluxetable*}{cccc}[!ht]
\digitalasset
\tablewidth{0pt}
\tablecaption{Diagnostic categories distribution for PSR B1933+16 (2019) \label{tab:class_stats}}
\tablehead{
\colhead{Category} & \colhead{Pulses} & \colhead{Class} & \colhead{Rate}
}
\startdata
(i) & 400 & QMP & 0.0382 \\
(ii) & 5,527 & NQMP & 0.5271 \\
(iii) & 4,175 & NQMP & 0.3981 \\
(iv) & 384 & NQMP & 0.0366 \\
\hline
Total & 10,486 & -- & 1.0000 \\
\enddata
\tablecomments{The four categories correspond exactly to the four types of single pulses illustrated in Figure~\ref{fig:diagnostic}. QMP = Quasi-periodic Modulation Pulse, NQMP = Non-Quasi-periodic Modulation Pulse.}
\end{deluxetable*}


 \begin{figure*}[ht!]
  \centering
  \captionsetup[subfigure]{labelformat=empty} 
  \renewcommand{\thesubfigure}{} 

  \begin{subfigure}[b]{0.30\textwidth}
    \centering
    \includegraphics[width=\textwidth]{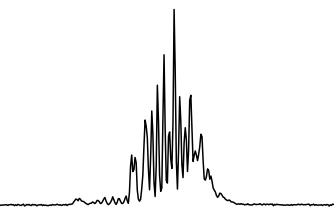}
    \caption{} 
  \end{subfigure}
  \hfill
  \begin{subfigure}[b]{0.30\textwidth}
    \centering
    \includegraphics[width=\textwidth]{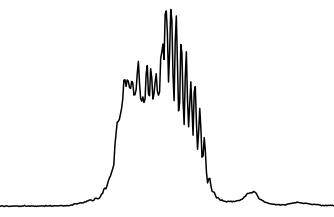}
    \caption{}
  \end{subfigure}
  \hfill
  \begin{subfigure}[b]{0.30\textwidth}
    \centering
    \includegraphics[width=\textwidth]{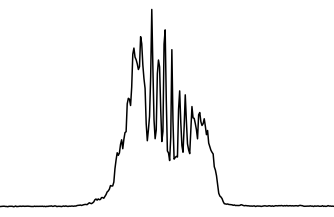}
    \caption{}
  \end{subfigure}

  \begin{subfigure}[b]{0.30\textwidth}
    \centering
    \includegraphics[width=\textwidth]{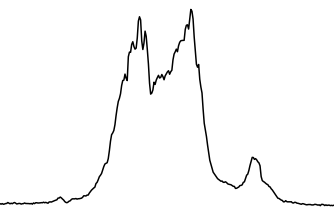}
    \caption{}
  \end{subfigure}
  \hfill
  \begin{subfigure}[b]{0.30\textwidth}
    \centering
    \includegraphics[width=\textwidth]{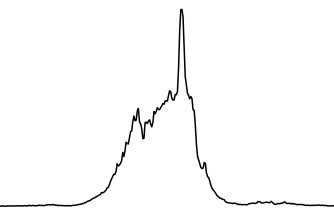}
    \caption{}
  \end{subfigure}
  \hfill
  \begin{subfigure}[b]{0.30\textwidth}
    \centering
    \includegraphics[width=\textwidth]{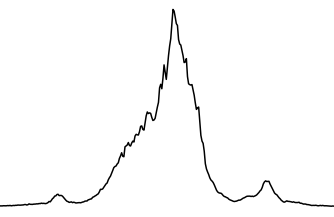}
    \caption{}
  \end{subfigure}

  \caption{Representative examples of positive and negative samples for single-pulse images. Top row: three positive samples (QMPs); bottom row: three negative samples (NQMPs).}
  \label{fig:sp}
\end{figure*}

\begin{figure*}[ht!]
  \centering
  \captionsetup[subfigure]{labelformat=empty} 
  \renewcommand{\thesubfigure}{} 

  \begin{subfigure}[b]{0.30\textwidth}
    \centering
    \includegraphics[width=\textwidth]{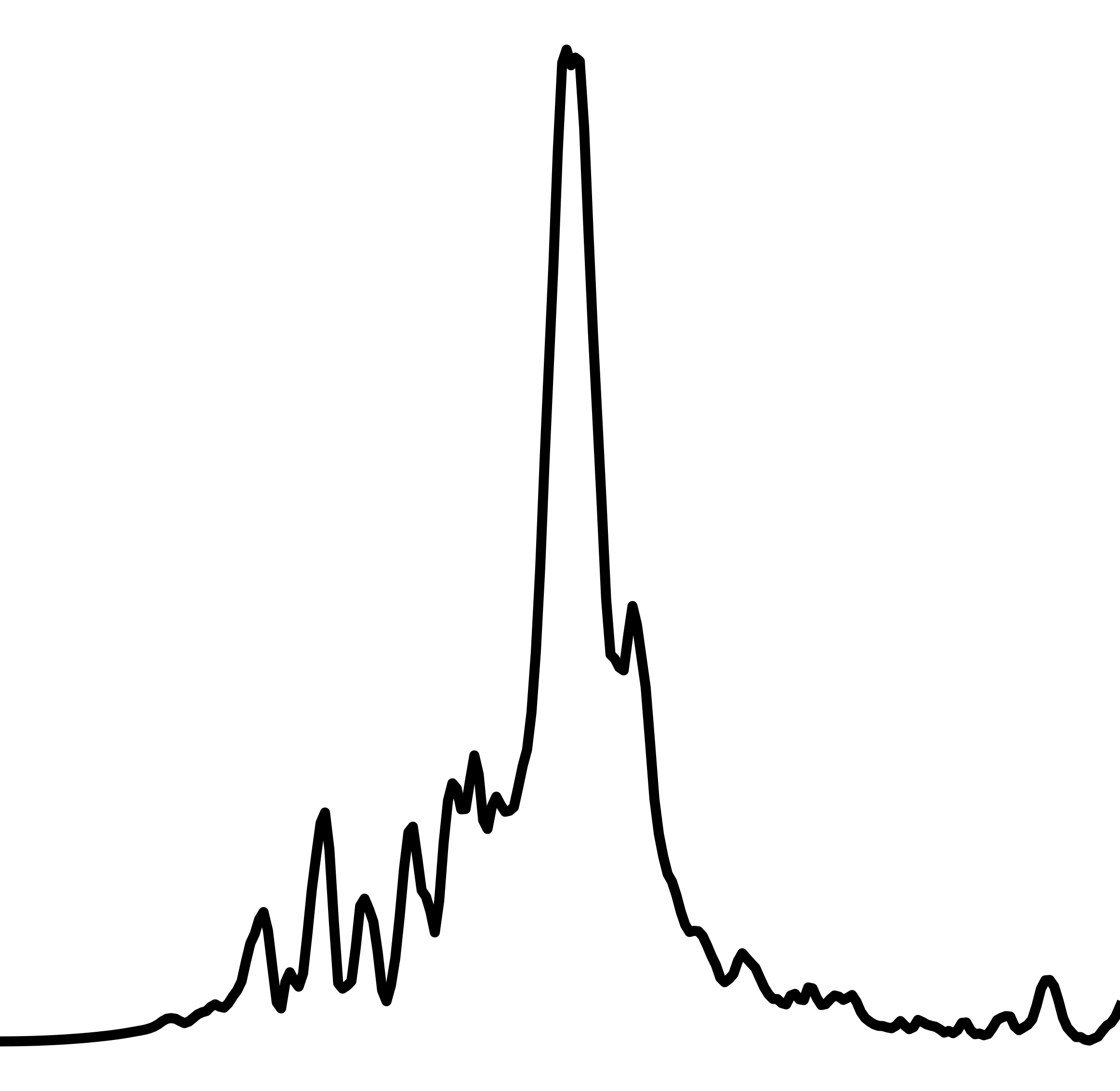}
    \caption{} 
  \end{subfigure}
  \hfill
  \begin{subfigure}[b]{0.30\textwidth}
    \centering
    \includegraphics[width=\textwidth]{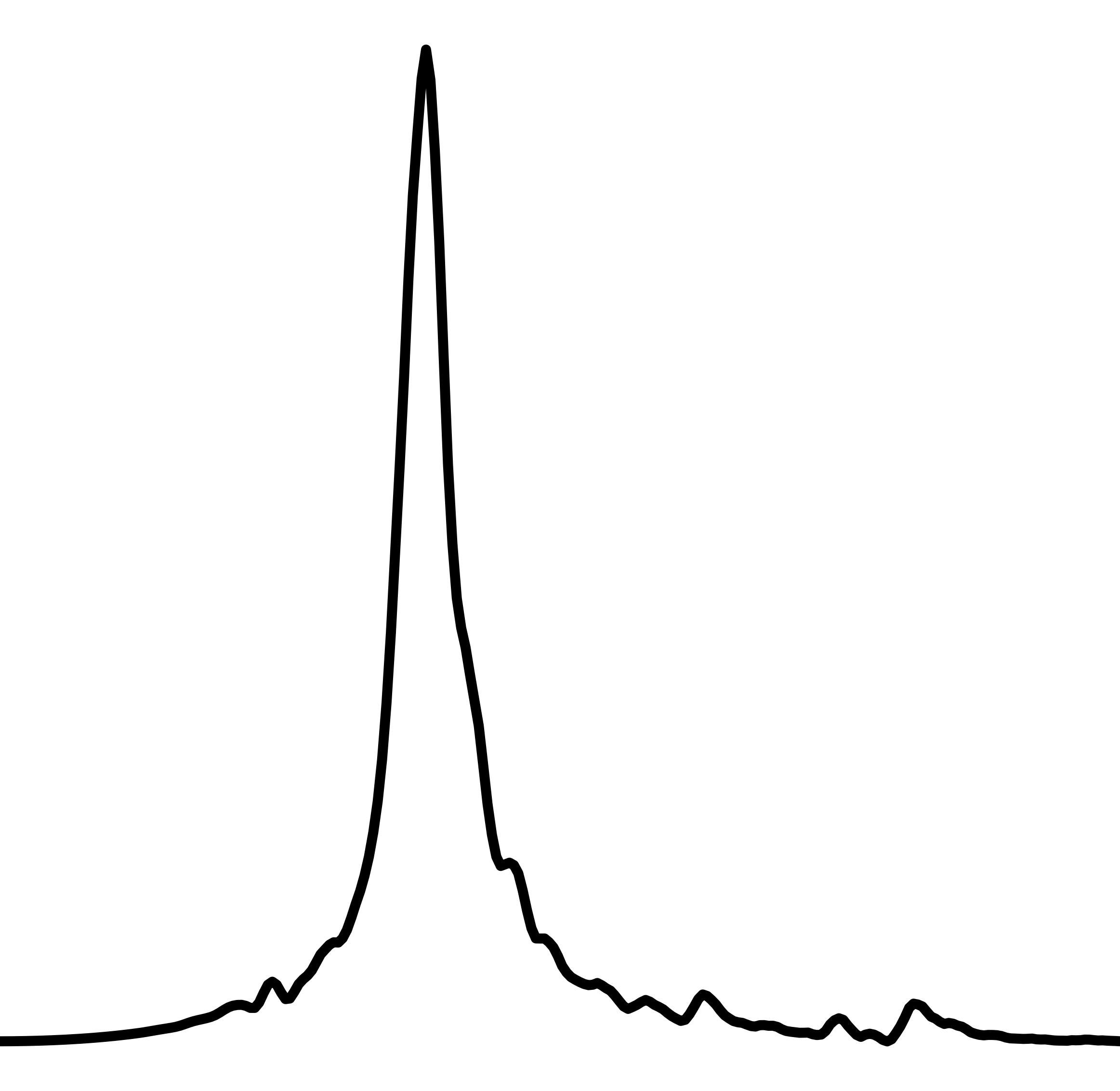}
    \caption{}
  \end{subfigure}
  \hfill
  \begin{subfigure}[b]{0.30\textwidth}
    \centering
    \includegraphics[width=\textwidth]{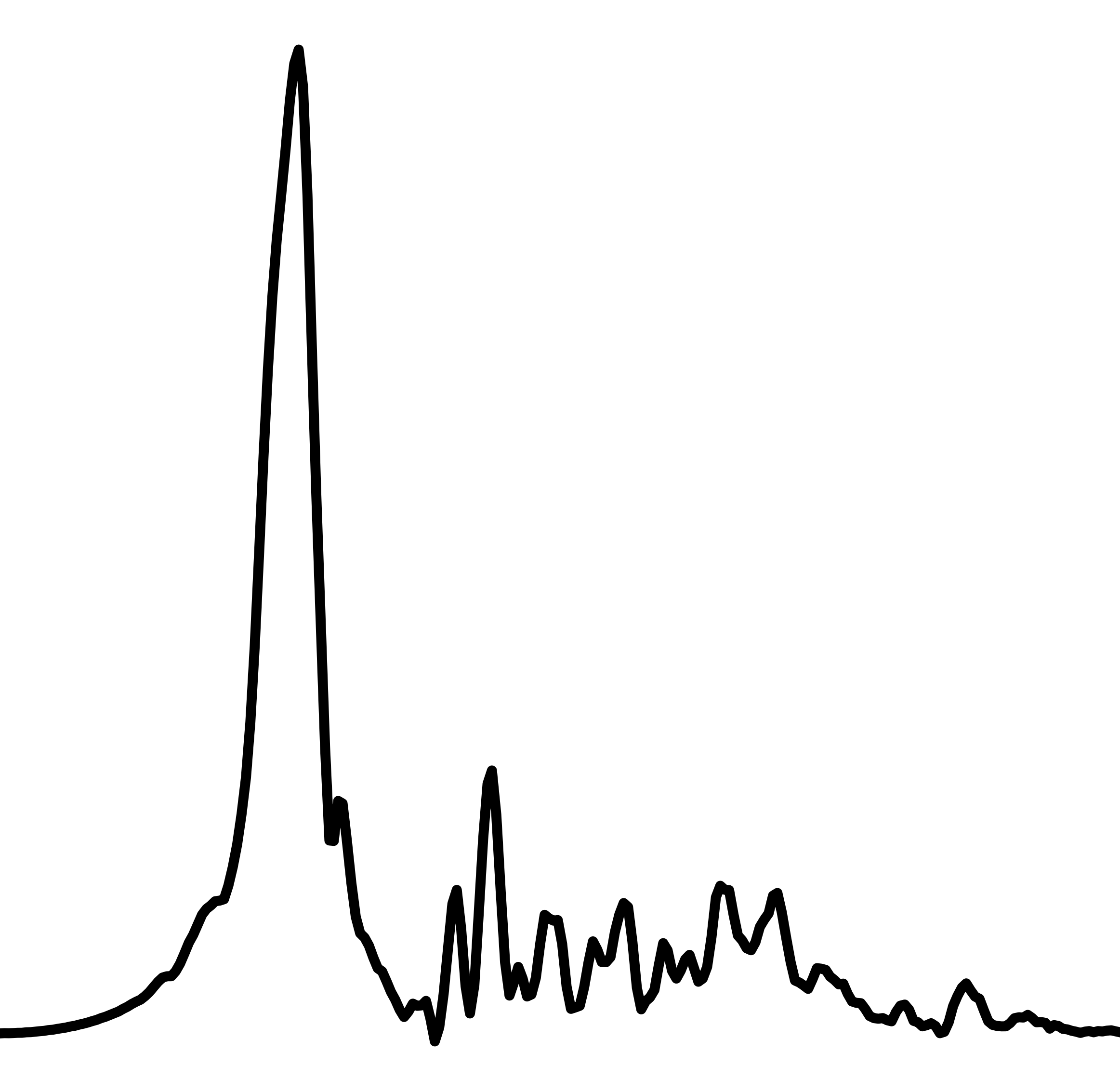}
    \caption{}
  \end{subfigure}

  \begin{subfigure}[b]{0.30\textwidth}
    \centering
    \includegraphics[width=\textwidth]{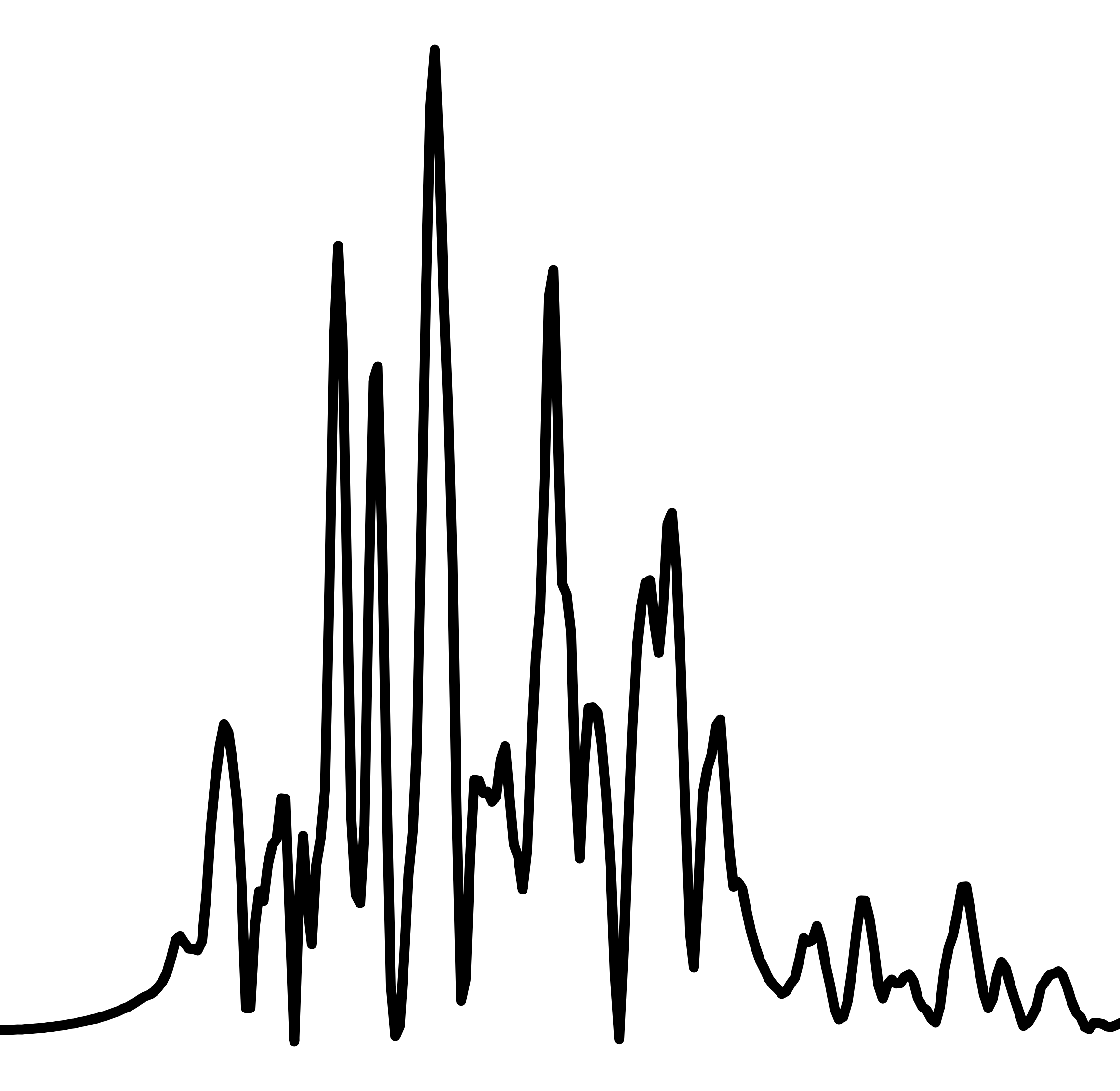}
    \caption{}
  \end{subfigure}
  \hfill
  \begin{subfigure}[b]{0.30\textwidth}
    \centering
    \includegraphics[width=\textwidth]{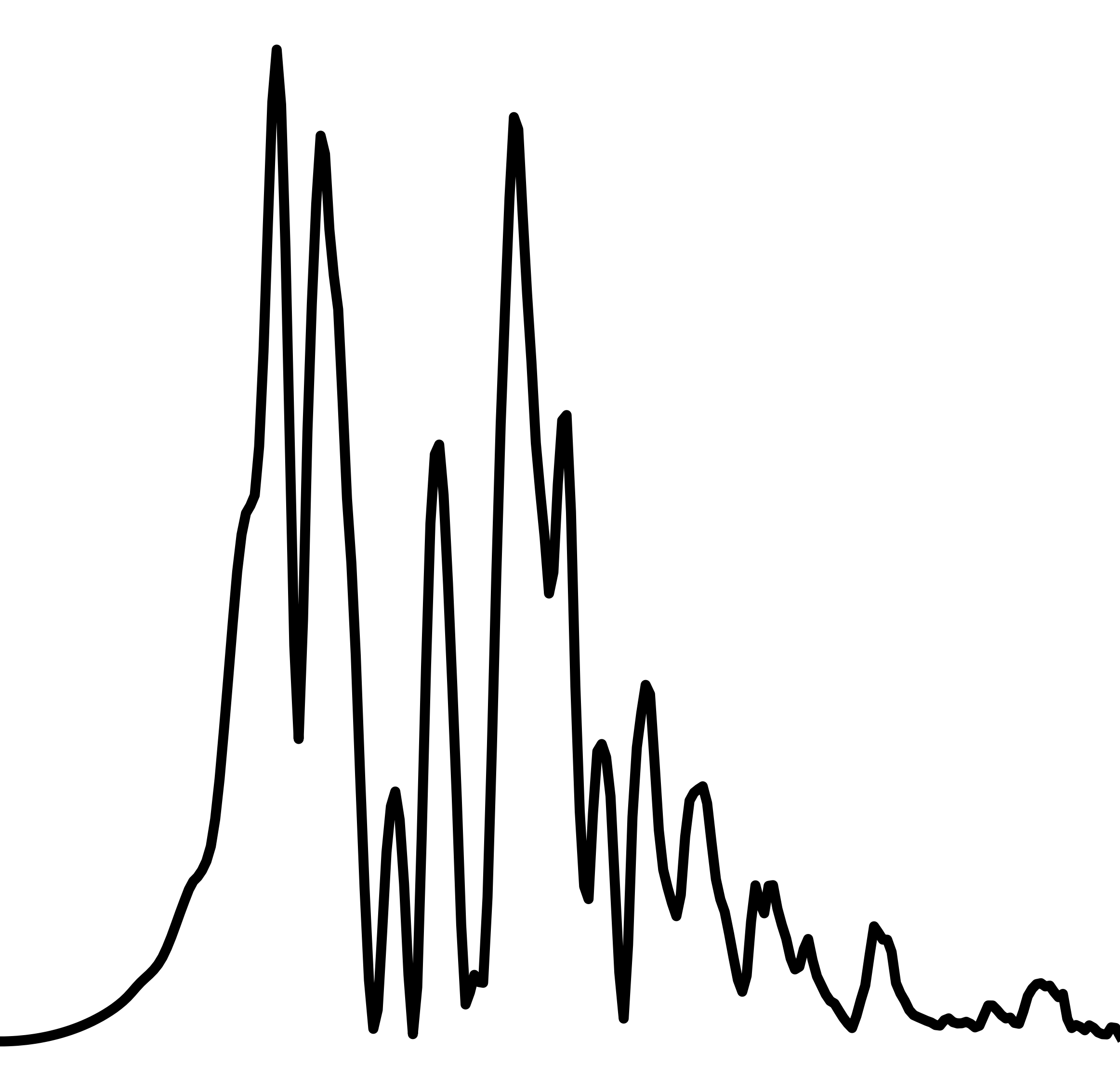}
    \caption{}
  \end{subfigure}
  \hfill
  \begin{subfigure}[b]{0.30\textwidth}
    \centering
    \includegraphics[width=\textwidth]{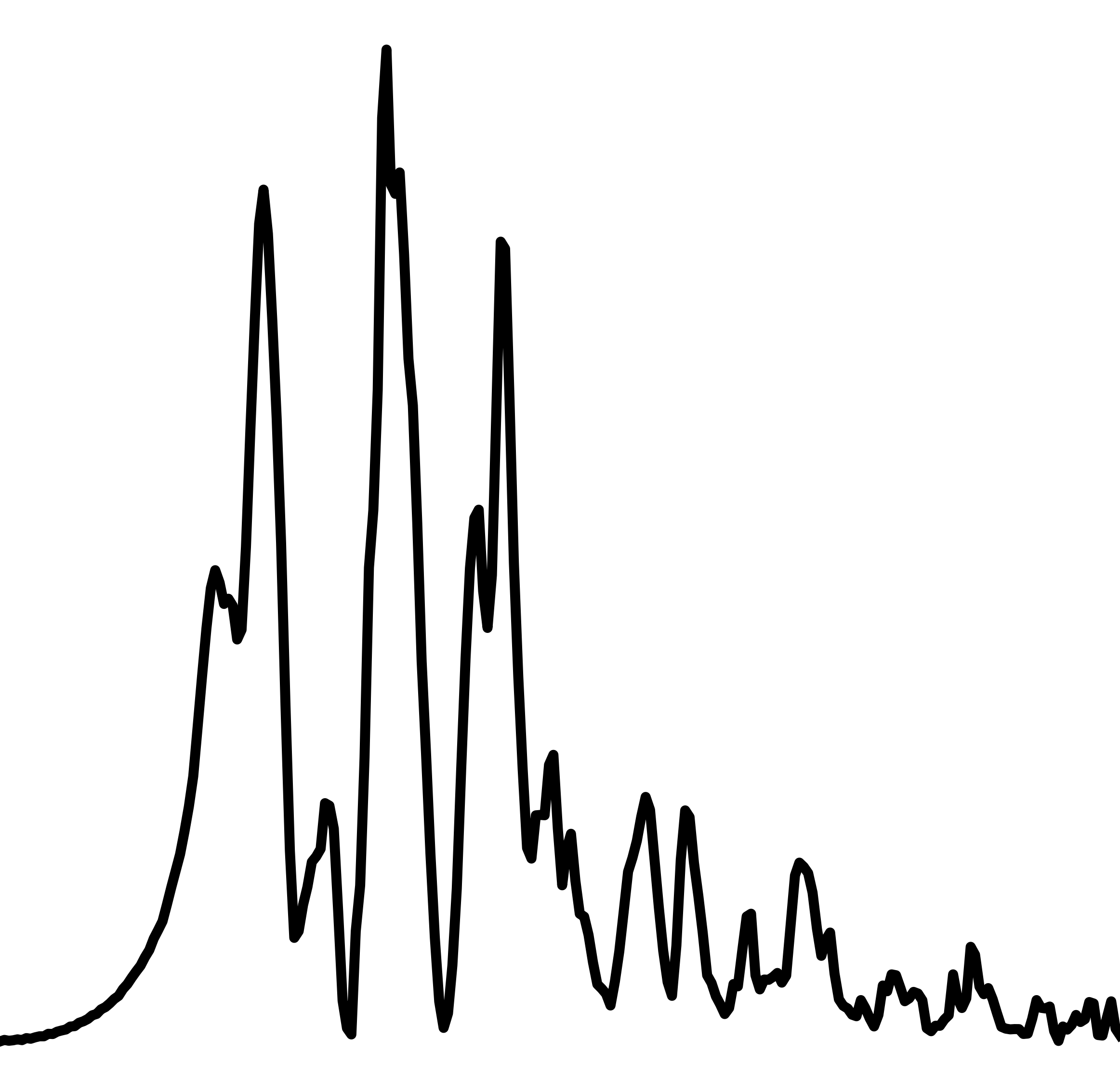}
    \caption{}
  \end{subfigure}

  \caption{Characteristic examples of positive and negative samples for ADP-derived power spectra. 
Top row: three positive samples (QMPs) exhibiting distinct unimodal signatures; 
bottom row: three negative samples (NQMPs) showing featureless spectra.}
  \label{fig:adp}
\end{figure*}

\subsection{Image Generation and Preprocessing}
\label{sec:Preprocessing}
Based on the complementary characteristics between single-pulse profiles and ADP spectra, we extract coordinate-free diagnostic images (specifically including single-pulse profiles (Figure~\ref{fig:sp}) and ADP spectra (Figure~\ref{fig:adp})) as model inputs, and maintained a balanced number of positive and negative samples for both image types (400 each, with 80\% for training and 20\% for validation), to enable the model to effectively learn intrinsic image features and stabilize training. 
Single-pulse and ADP images are determined by the configuration of the data generation and visualization code, measuring $334\times217$ and $2325\times2265$ pixels respectively. 
All input images underwent an identical preprocessing pipeline: they were first resized to a uniform 224$\times$224 resolution. The resulting image was then converted to RGB color format \footnote{This format conversion was performed only to maintain compatibility with the standard ResNet architecture.} and subjected to Z-score normalization using specified mean and standard deviation values, before finally being packaged for model input.

\section{Dual-Stage Residual Network Model} \label{sec:dsr-model}

\begin{figure}[htbp]
\centering
\includegraphics[width=0.95\textwidth]{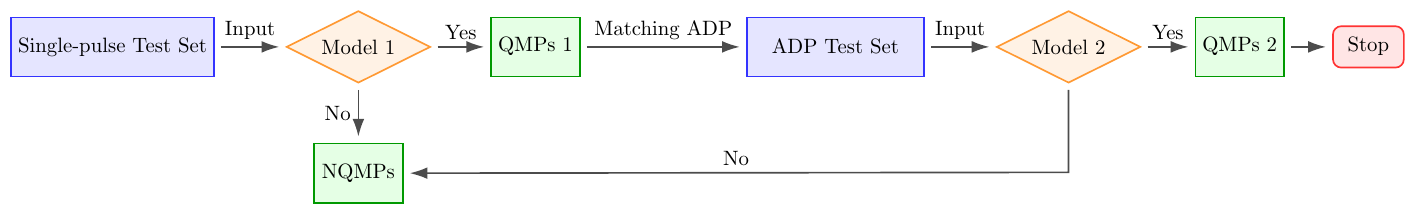}
\caption{Schematic diagram of QMP identification by the DSR system. Model 1 and Model 2 serve as conditional judgment mechanisms that determine classification by comparing whether an image's positive-class prediction probability exceeds its negative-class probability: if yes, it is classified as positive; otherwise as negative. The input consists of single-pulse test images, and the final output comprises ADP images matched with QMPs.}
\label{fig:dsr_testing}
\end{figure}

Convolutional Neural Networks (CNN) effectively extract spatial features and deliver superior performance in image classification tasks \citep{LeCun2015}, with Residual Networks (ResNets) further enhancing model performance (\citet{He2016}; \citet{Hardt2016}; \citet{DU2018arXiv180600900D}). 
Our analysis demonstrates that reliable QMP identification requires joint analysis of both single-pulse images and ADP plots (Figure~\ref{fig:diagnostic}), as single ResNets processing unimodal images yield high false-positive rates; 
consequently, we developed a DSR-based QMP identification workflow (Figure~\ref{fig:dsr_testing}) that synergistically fuses features from single-pulse and ADP images: single-pulse inputs are first processed by Model 1, where signals identified as candidates are matched by pulse ID with pre-generated ADP images and passed to Model 2. Only signals authenticated by both models are confirmed as genuine QMPs. 

\begin{figure}[htbp]
\centering
\begin{subfigure}[b]{0.85\textwidth}
  \centering
  \includegraphics[width=\textwidth]{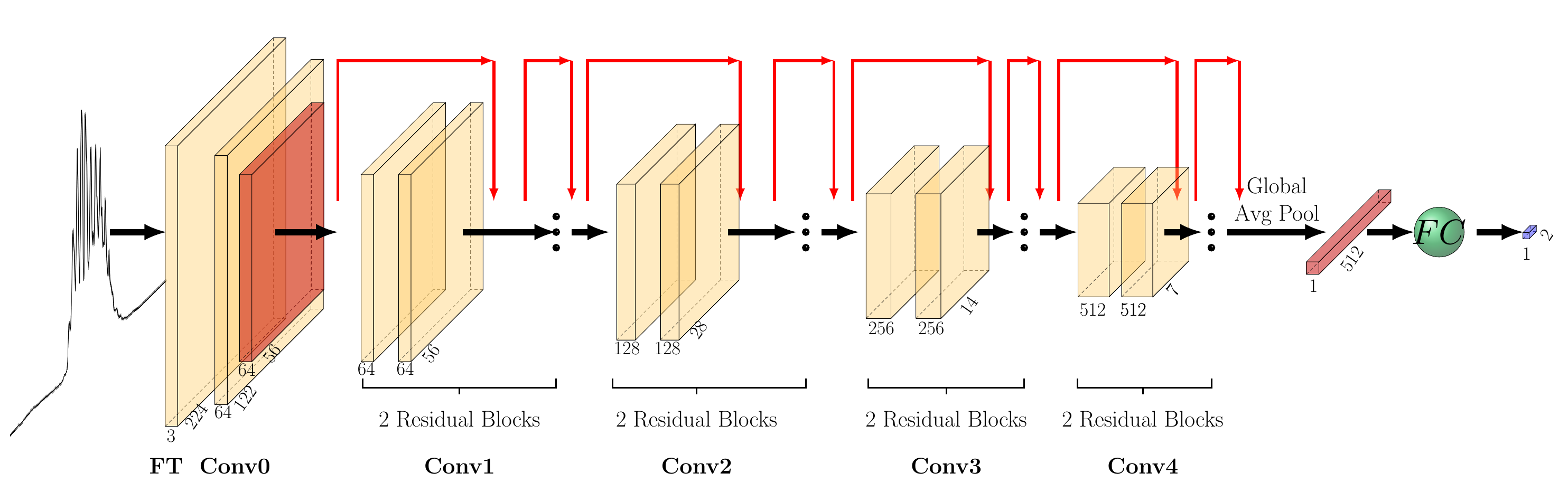}
  \caption{ResNet-18 architecture for single-pulse images}
  \label{fig:resnet18}
\end{subfigure}
\vspace{1em} 
\begin{subfigure}[b]{0.85\textwidth}
  \centering
  \includegraphics[width=\textwidth]{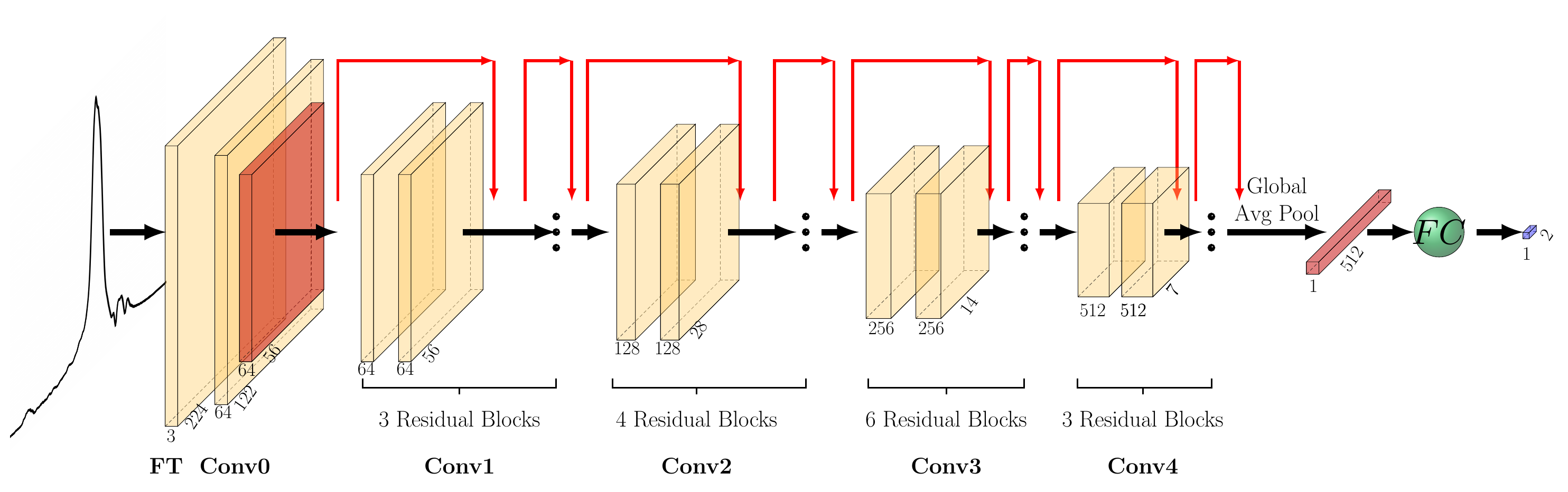}
  \caption{ResNet-34 architecture for ADP images}
  \label{fig:resnet34}
\end{subfigure}
\caption{Workflow of the DSR model: 
(a) ResNet-18 processes single-pulse images; 
(b) ResNet-34 processes ADP images. 
Common components: initial convolutional layer (yellow), residual block groups, global average pooling (brown bar), fully connected layer with Softmax activation (green sphere), and generation of 2D arrays (blue blocks). 
For ResNet-18, each residual block group contains 2 residual blocks; ResNet-34 contains 3, 4, 6, and 3 blocks per group respectively.}

\label{fig:dsr_workflow}
\end{figure}

\begin{figure}[htbp]
\centering
\begin{subfigure}[b]{0.49\textwidth}
  \centering
  \includegraphics[width=\textwidth]{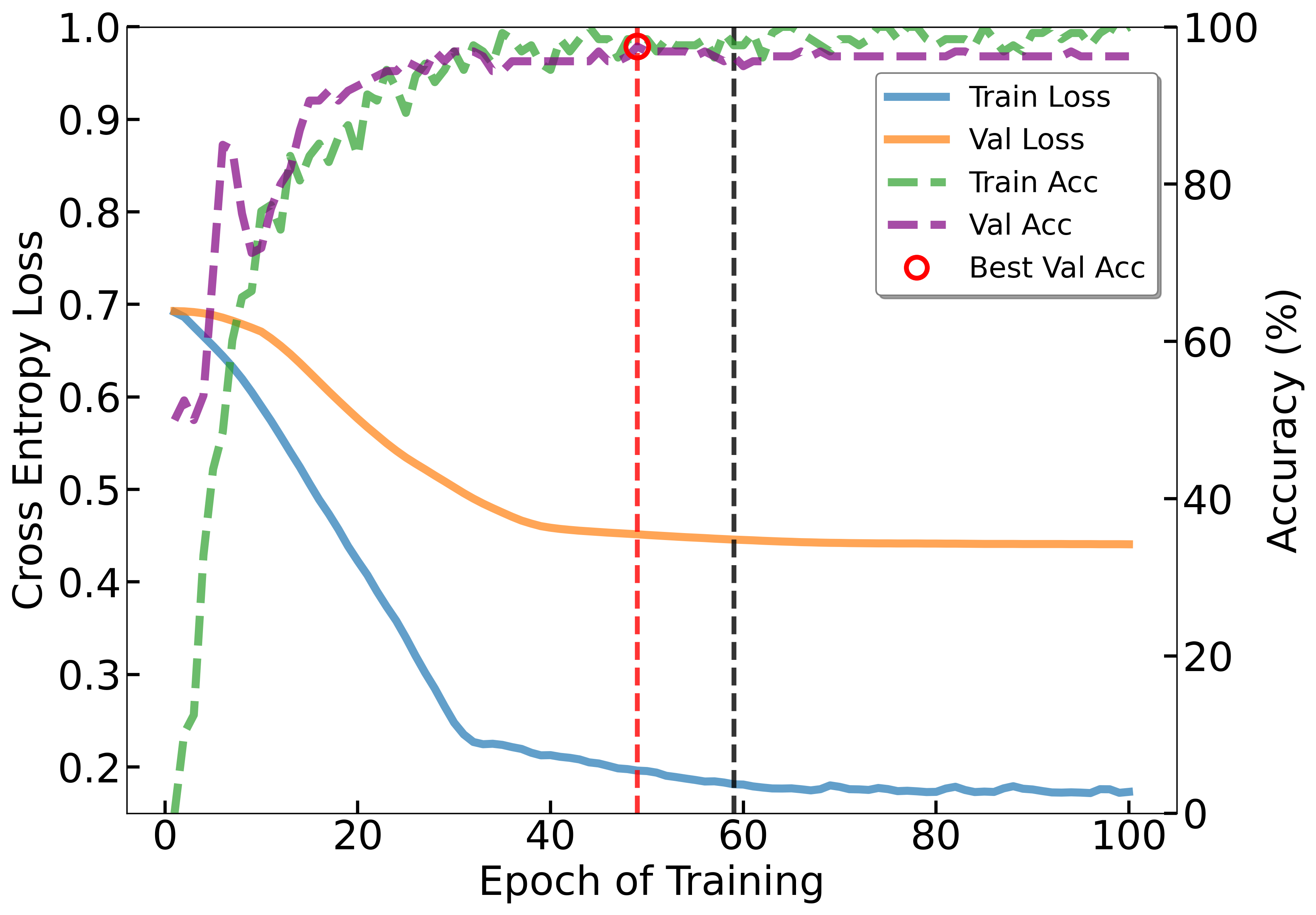}
  \caption{ResNet-18: Single-pulse images}
  \label{fig:resnet18_training}
\end{subfigure}
\hfill
\begin{subfigure}[b]{0.49\textwidth}
  \centering
  \includegraphics[width=\textwidth]{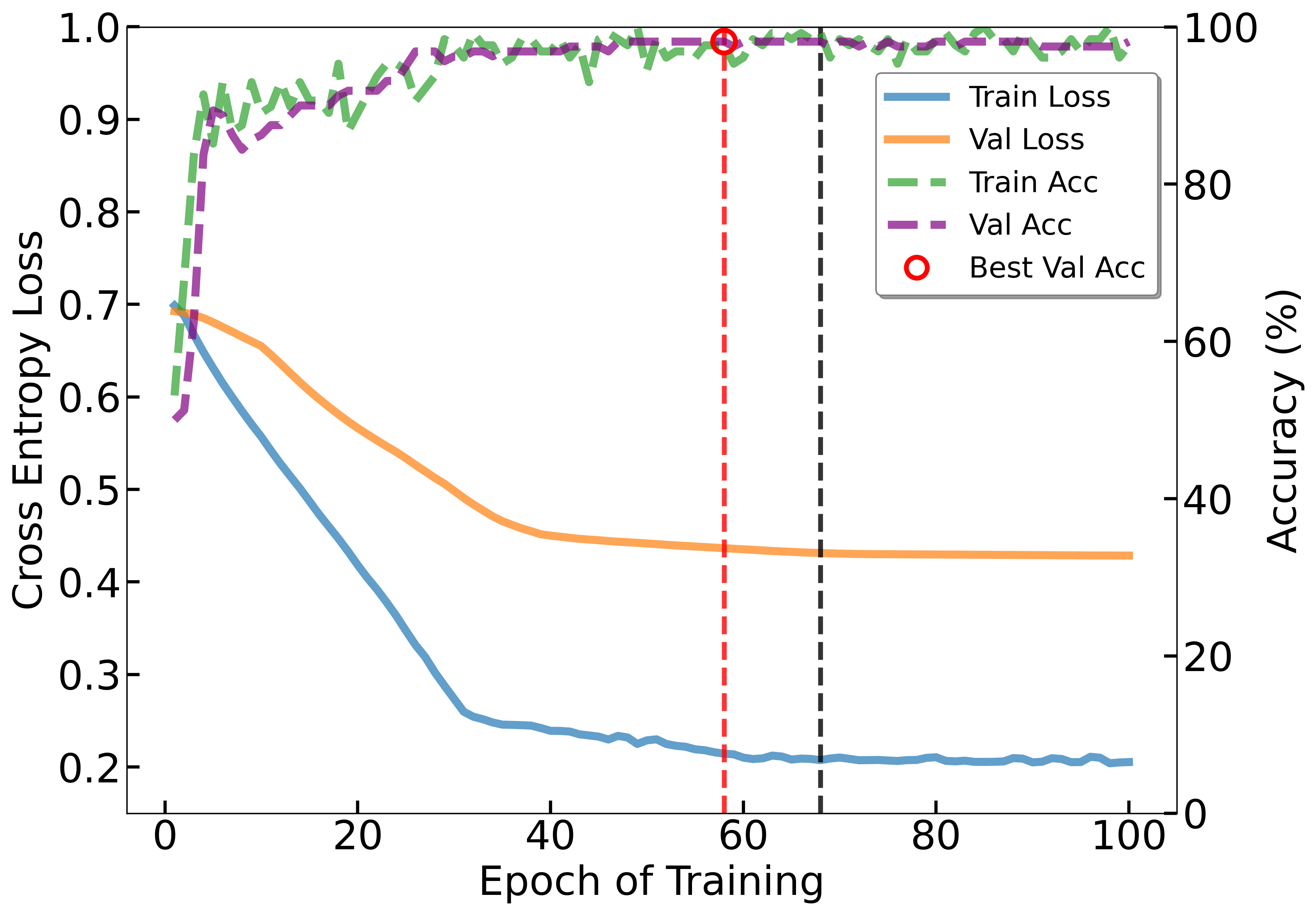}
  \caption{ResNet-34: ADP images}
  \label{fig:resnet34_training}
\end{subfigure}

\caption{Evolution of accuracy and loss during dual-stage ResNet training. 
(a) Green and purple dashed lines: training and validation accuracy; 
blue and orange solid lines: training and validation loss. 
Vertical red dashed line: optimal performance point (epoch 49); 
black dashed line: early stopping point (epoch 59); 
red hollow circle: peak validation accuracy (97.50\%). 
(b) Corresponding curves for ResNet-34 with optimal point at epoch 58 (98.13\% accuracy) and early stopping at epoch 68.}
\label{fig:training_curves}
\end{figure}

To achieve optimal model performance, we conducted comparative tests using ResNets of varying depths, ultimately determining that Model 1 is generated by training ResNet-18 on single-pulse images while Model 2 is generated by training ResNet-34 on ADP images (detailed in Section~\ref{sec:test}). Both architectures share fundamental components (Figure~\ref{fig:dsr_workflow}): an initial convolutional layer (yellow), grouped residual blocks, global average pooling (brown bar), a fully-connected layer with Softmax activation (green sphere), and a 2D feature output (blue blocks). However, their residual architectures differ significantly: Model 1 (ResNet-18) contains 2 residual blocks per group, whereas Model 2 (ResNet-34) contains 3, 4, 6, and 3 blocks per group respectively. Following the dataset partitioning scheme in Section~\ref{sec:Preprocessing}, both image types maintain independent training, validation, and test sets. For single-pulse image processing, ResNet-18 underwent 100-epoch training , with post-epoch evaluations on both sets enabling dynamic hyperparameter adjustments until loss and accuracy metrics demonstrated favorable convergence trends (Figure~\ref{fig:resnet18_training}). Final hyperparameters were set as: training batch size=128, validation batch size=32, AdamW optimizer, learning rate=8.6e-5, and weight decay=0.02. Subsequently, ResNet-34 was trained on ADP images with a learning rate of 5.7e-5, with the same batch sizes, optimizer, and weight decay settings (Figure~\ref{fig:resnet34_training}). To prevent overfitting, we employed an early stopping strategy \citep{Zeng2020} and identified the optimal performance points within the early stopping range (indicated by red hollow circles in Figure~\ref{fig:training_curves}). ResNet-18 and ResNet-34 achieved their best performance at Epoch 49 and Epoch 58 respectively, at which point the models were saved as Model 1 and Model 2 in Figure~\ref{fig:dsr_testing}.

\section{Experimental Results and Analysis}  \label{sec:re}
\subsection{Performance Evaluation Metrics} \label{sec:4.1}
To comprehensively assess model performance, we employ accuracy, recall, precision, and F1 score as primary evaluation metrics \citep{Lin2020}. In this binary classification task: Accuracy represents the proportion of correctly predicted samples out of the total predictions made by a model. Recall measures the proportion of true QMPs correctly identified, reflecting the model's detection completeness. Precision quantifies the proportion of true QMPs among all candidates classified as QMPs, indicating detection accuracy. F1 score represents the harmonic mean of recall and precision, providing a balanced evaluation of overall performance. The formal definitions are:  

\begin{align}
\text{Accuracy} &= \frac{\text{TP} + \text{TN}}{\text{TP} + \text{FN} + \text{FP} + \text{TN}} \\
\text{Precision} &= \frac{\text{TP}}{\text{TP} + \text{FP}} \\
\text{Recall} &= \frac{\text{TP}}{\text{TP} + \text{FN}} \\
\text{F1 score} &= \frac{2 \times \text{Precision} \times \text{Recall}}{\text{Precision} + \text{Recall}}
\end{align}  

where: TP (True Positives): Correctly identified QMPs. FP (False Positives): NQMPs misclassified as QMPs. FN (False Negatives): Undetected QMPs. TN (True Negatives): Correctly rejected NQMPs

\subsection{Test Results}  \label{sec:test}
Based on the performance comparisons in Table~\ref{tab:resnet_combined}, ResNet-18 and ResNet-34 were identified as the optimal architectures for single-pulse and ADP image detection, respectively. Consequently, the DSR framework employs ResNet-18 in Stage 1 (yielding Model 1) and ResNet-34 in Stage 2 (yielding Model 2).
A brief analysis of the other architectures reveals a general trend: deeper networks (ResNet-50 and ResNet-101) do not necessarily yield better performance on this specific task. For single-pulse detection, while ResNet-50 achieves perfect precision (1.0000), its recall is the lowest (0.9000), indicating a tendency towards more conservative predictions. ResNet-101 shows a slight improvement in recall over ResNet-50 but still underperforms compared to the shallower ResNet-18 and ResNet-34. A similar pattern was observed in ADP detection, where the deepest model ResNet-101 performed competently overall but was still marginally surpassed by the shallower ResNet-34. This suggests that for the given dataset, the optimal model complexity is achieved at a moderate depth, avoiding over-parameterization and potential overfitting.

\begin{deluxetable*}{ccccccc}[!ht]
\digitalasset
\tablewidth{0pt}
\tablecaption{Performance Comparison of Different ResNet Architectures on Single-Pulse and ADP Images from PSR B1933+16 Validation Set
\label{tab:resnet_combined}}
\tablehead{
\colhead{Input Type} & \colhead{ResNet} & \colhead{Optimal Epoch} & \colhead{Accuracy} & \colhead{Precision} & \colhead{Recall} & \colhead{F1-score}
}
\startdata
Single-Pulse Images & ResNet-18 & 49 & 0.9750 & 0.9871 & 0.9625 & 0.9747 \\
...& ResNet-34 & 59 & 0.9625 & 0.9744 & 0.9500 & 0.9620 \\
...& ResNet-50 & 30 & 0.9500 & 1.0000 & 0.9000 & 0.9473 \\
...& ResNet-101 & 14 & 0.9563 & 0.9867 & 0.9250 & 0.9548 \\
\hline
ADP Images & ResNet-18 & 60 & 0.9563 & 0.9195 & 1.0000 & 0.9580 \\
...& ResNet-34 & 58 & 0.9813 & 0.9639 & 1.0000 & 0.9816 \\
...& ResNet-50 & 24 & 0.9563 & 0.9294 & 0.9875 & 0.9575 \\
...& ResNet-101 & 19 & 0.9750 & 0.9634 & 0.9875 & 0.9753 \\
\enddata
\tablecomments{The optimal epochs were determined using early stopping. Performance metrics compare different ResNet architectures on the validation set of PSR B1933+16 for QMP detection. All metrics are reported with 4-digit precision. While all compared ResNet architectures follow the same macroscopic structure shown in Figure \ref{fig:dsr_workflow}, ResNet-50 and ResNet-101 employ bottleneck residual blocks (with three convolutional layers per block) rather than the basic blocks (with two convolutional layers) used in ResNet-18 and ResNet-34. This architectural difference, in addition to the increased number of blocks, accounts for their greater depth and parameter count.}
\end{deluxetable*}

\begin{figure}[htbp]
\centering
\includegraphics[width=0.80\textwidth]{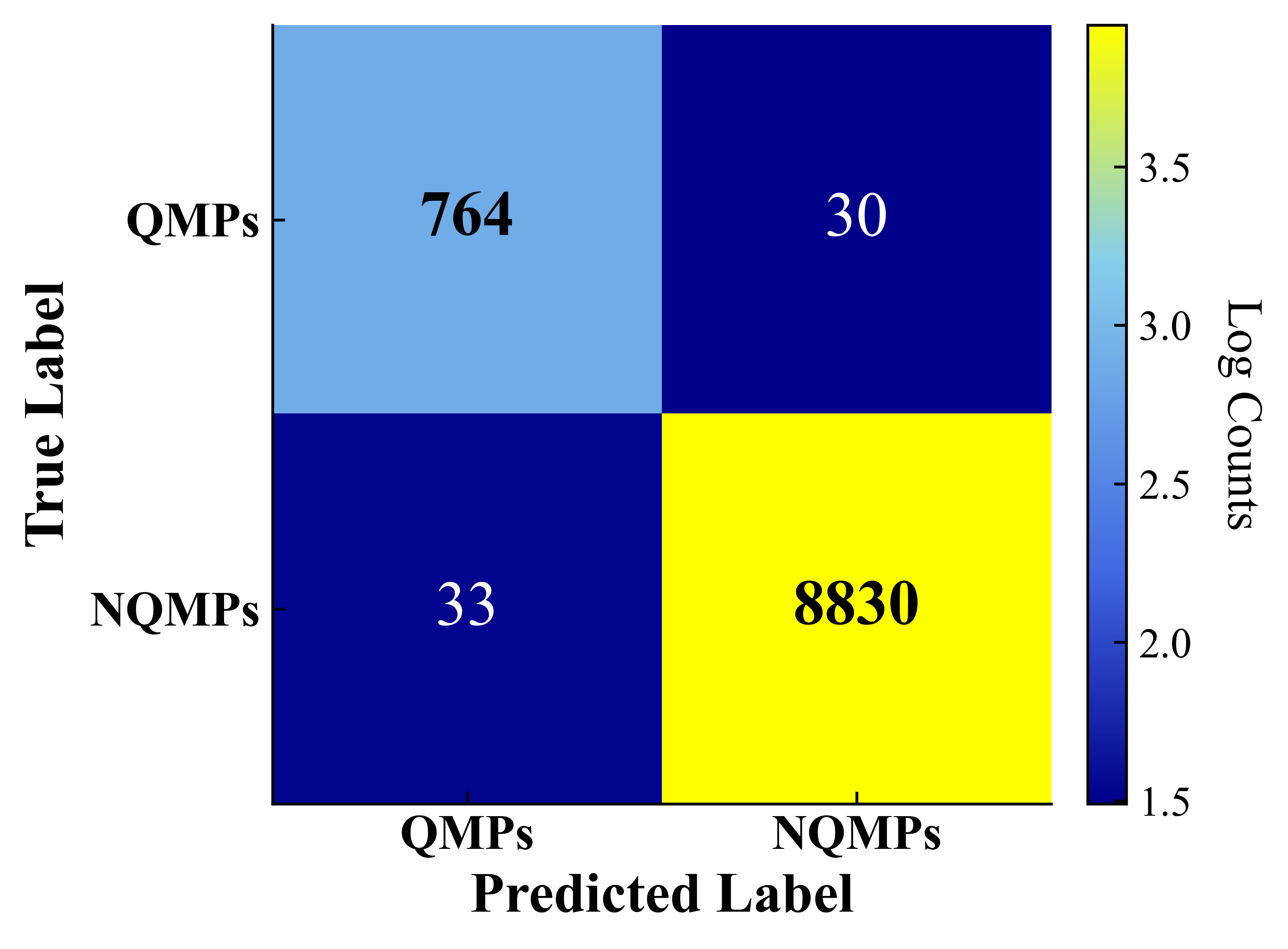} 
\caption{
Confusion matrix for ResNet testing on PSR B1933+16 dataset (using DSR model on 2020 data). 
The x-axis represents predicted labels, while the y-axis represents true labels. 
The color bar indicates the distribution of the logarithmic transformation of TP, FP, TN, and FN counts. 
Specific numerical values: TP=764, FP=33, TN=8830, FN=30.
}
\label{fig:confusion_matrix_example}
\end{figure}

Executing QMP identification on the PSR B1933+16 (2020) test set with these trained models, we constructed confusion matrices (Figure~\ref{fig:confusion_matrix_example}) and calculated performance metrics (Table~\ref{tab:performance_comparison}), revealing that Model 1 alone achieved precision = 0.2193 and Model 2 precision = 0.5236, whereas the integrated DSR approach significantly elevated precision to 0.9585 (recall = 0.9610), substantially outperforming single-stage models (detailed analysis in Section~\ref{subsec:single_stage_analysis}). For cross-year and cross-pulsar validation, we employed a post-hoc verification protocol where candidates were confirmed as genuine QMPs only when simultaneously satisfying two diagnostic criteria: exhibiting pronounced clustered microstructure in single-pulse profiles (Figure~\ref{fig:sub1}a) and demonstrating unimodal distribution characteristics in ADP plots (Figure~\ref{fig:sub1}d) with SNR \textgreater 10 (Section~\ref{sec:classification}). The 2021 dataset for PSR B1933+16 yielded 906 candidates with 862 verified genuine QMPs (precision=0.9516), while the 2022 dataset produced 979 candidates with 942 confirmed genuine QMPs (precision=0.9617). Cross-pulsar validation results showed: 33 genuine QMPs among 35 candidates for PSR J0034-0721 (precision=0.9459); 212 genuine QMPs among 228 candidates from the DC component of PSR J0304+1932 (precision=0.9306); and 34 genuine QMPs among 36 candidates from its LC component (precision=0.9473). These experiments demonstrate that the DSR framework maintains exceptional performance across multi-year, multi-pulsar, and multi-emission-component tests (precision consistently \textgreater 0.93), systematically validating its generalization capability for efficient QMP detection in large-scale sky survey data.


\begin{deluxetable*}{cccccccc}[!ht]
\digitalasset
\tablewidth{0pt}
\tablecaption{Performance Comparison of Different Models and DSR Test Results on Cross-Year and Cross-Pulsar Datasets \label{tab:performance_comparison}}
\tablehead{
\colhead{Source} & \colhead{Total Pulses} & \colhead{Model} & \colhead{Precision} &\colhead{Recall} &\colhead{F1-score} &\colhead{TP} &\colhead{FN}
}
\startdata
PSR B1933+16 (2020) & 9657 & Model 1  & 0.2193 & 1.0000 & 0.3598 & 764 & 0 \\
PSR B1933+16 (2020) & 9657 & Model 2  & 0.5236 & 0.9610 & 0.6780 & 764 & 30 \\
PSR B1933+16 (2020) & 9657 & DSR (Model 1+2) & 0.9585 & 0.9610 & 0.9597 & 764 & 30 \\
PSR B1933+16 (2021) & 9704 & DSR (Model 1+2) & 0.9516 & -- & -- & 906 & -- \\
PSR B1933+16 (2022) & 9752 & DSR (Model 1+2) & 0.9617 & -- & -- & 979 & -- \\ 
PSR J0034-0721 & 3118 & DSR (Model 1+2) & 0.9459 & -- & -- & 35 & -- \\ 
PSR J0304+1932 (DC) & 3414 & DSR (Model 1+2) & 0.9306 & -- & -- & 228 & -- \\ 
PSR J0304+1932 (LC) & 3414 & DSR (Model 1+2) & 0.9473 & -- & -- & 36 & -- \\ 
\enddata
\tablecomments{Model 1 processes single-pulse images. Model 2 analyzes ADP images. DSR represents the dual-stage combination. TP = True Positives (detected QMPs), FN = False Negatives (missed QMPs). Accuracy holds limited significance for evaluating QMP detection performance in scenarios where negative samples substantially outnumber positive samples, and is therefore not considered here. Recall, F1-score and FN not calculated for 2021-2022 datasets and cross-pulsar data (marked with ``--").}
\end{deluxetable*}

\section{Discussion}\label{sec:dis}


\subsection{Comparison of 1D Data and Images as Model Input}

The approach of employing a 1D CNN to process the three raw sequences, namely single-pulse profile, PSD, and ADP, demonstrates potential advantages in training efficiency. To systematically evaluate the performance of this alternative approach and validate the rationale behind our adopted 2D image-based methodology, We constructed a three-stream 1D CNN as a baseline model, and performed model training and comparative testing analysis by inputting the one-dimensional raw sequences homologous to the image data.

The architecture of this baseline model is designed as follows: three independent 1D sequences (each with data dimensions of $N \times 1 \times L$) are fed into dedicated Conv1d modules (with a kernel size of 7 and a stride of 2) for feature extraction. Each module outputs a feature tensor of dimensions $N \times 64 \times 512$. These feature tensors are then dimensionally expanded into a four-dimensional form of $N \times 1 \times 64 \times 512$, and subsequently concatenated and fused along the channel dimension to form a composite feature tensor of $N \times 3 \times 64 \times 512$. This composite tensor is finally input into standard ResNet architectures (including variants with 18, 34, 50, and 101 layers) to complete the classification task.

Comparative experimental results, as shown in Table~\ref{tab:performance_comparison1}, clearly reveal the performance trade-offs between the different input modalities: although the 1D ResNet variants significantly outperform our 2D DSR model in training efficiency (e.g., ResNet-18 requires only 201 seconds of training time), this gain in efficiency comes at the cost of a substantial reduction in core classification performance metrics, with precision being particularly affected. Our 2D DSR model achieved a precision of 0.9585 and an F1-score of 0.9597 on the PSR B1933+16 (2020) test set, whereas the best-performing 1D model (ResNet-34) attained corresponding metrics of only 0.4727 and 0.6097, indicating a significant performance gap. The decision to adopt image input was further supported by two additional considerations: first, the 2D visualization format aligns more closely with human expert cognitive patterns, effectively enhancing interpretability; second, the image-based approach offers greater adaptability to cross-instrument data variations, avoiding the complex preprocessing pipelines required for 1D data. Considering this performance advantage alongside these complementary benefits, we ultimately selected images as the input data for our model.

\begin{deluxetable*}{cccccccc}[!ht]
\digitalasset
\tablewidth{0pt}
\tablecaption{Performance Comparison of QMP Identification Using Input Data of Different Dimensionalities
\label{tab:performance_comparison1}}
\tablehead{
\colhead{Source} & \colhead{Total} & \colhead{Model} & \colhead{Precision} &\colhead{Recall} &\colhead{F1-score} &\colhead{Input} &\colhead{Training} \\
\colhead{} & \colhead{pulses} & \colhead{} & \colhead{} &\colhead{} &\colhead{} &\colhead{} &\colhead{time (s)}
}
\startdata
PSR B1933+16 (2020) & 9657 & DSR  & 0.9585 & 0.9610 & 0.9597 & Two 2D images & 5315 \\
PSR B1933+16 (2020) & 9657 & ResNet-18  & 0.4373 & 0.9143 & 0.5917 & Three 1D arrays & 201 \\
PSR B1933+16 (2020) & 9657 & ResNet-34  & 0.4727 & 0.8589 & 0.6097 & Three 1D arrays & 240\\
PSR B1933+16 (2020) & 9657 & ResNet-50  & 0.3710 & 0.8310 & 0.5128 & Three 1D arrays & 305\\
PSR B1933+16 (2020) & 9657 & ResNet-101  & 0.2272 & 0.8489 & 0.3585 & Three 1D arrays & 438 \\
\enddata
\tablecomments{DSR (2D) represents the dual-stage residual network processing two 2D images (single-pulse profile and ADP spectrum). 1D models process flattened versions of the same data as three 1D arrays. Training time represents the total training duration required for model convergence. Accuracy holds limited significance for evaluating QMP detection performance in scenarios where negative samples substantially outnumber positive samples, and is therefore not considered here.}
\end{deluxetable*}

\subsection{Analysis of Performance Discrepancies in Single-Stage Models}
\label{subsec:single_stage_analysis}

\begin{figure*}[ht!]
\centering
\captionsetup[subfigure]{labelformat=simple}
\renewcommand{\thesubfigure}{(\roman{subfigure})} 

\begin{subfigure}[b]{0.48\textwidth}
    \centering
    \includegraphics[width=\textwidth]{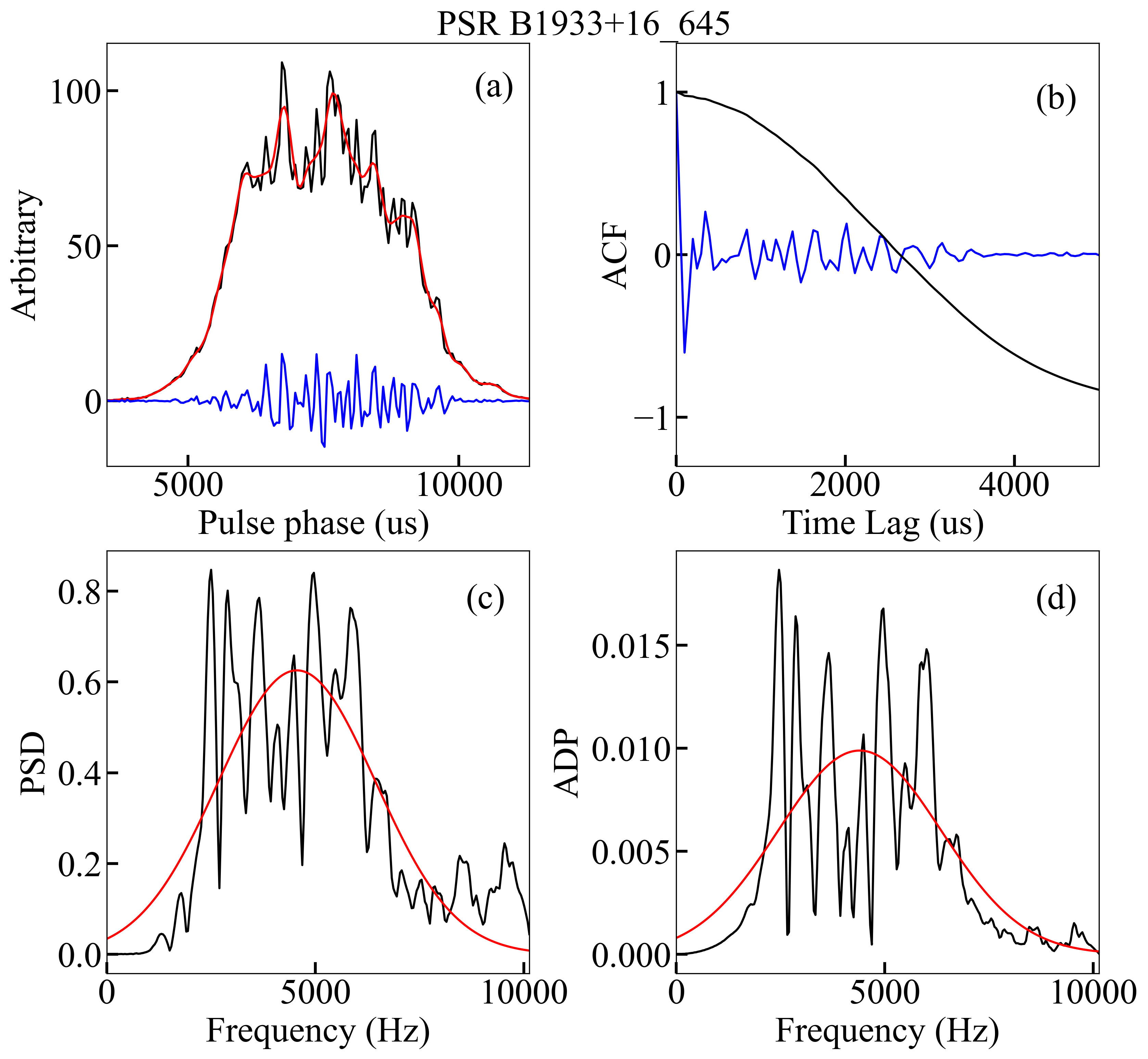}
    \caption{False positive sample 645 } 
    \label{fig:pseudo_qmp1}
\end{subfigure}
\hfill
\begin{subfigure}[b]{0.48\textwidth}
    \centering
    \includegraphics[width=\textwidth]{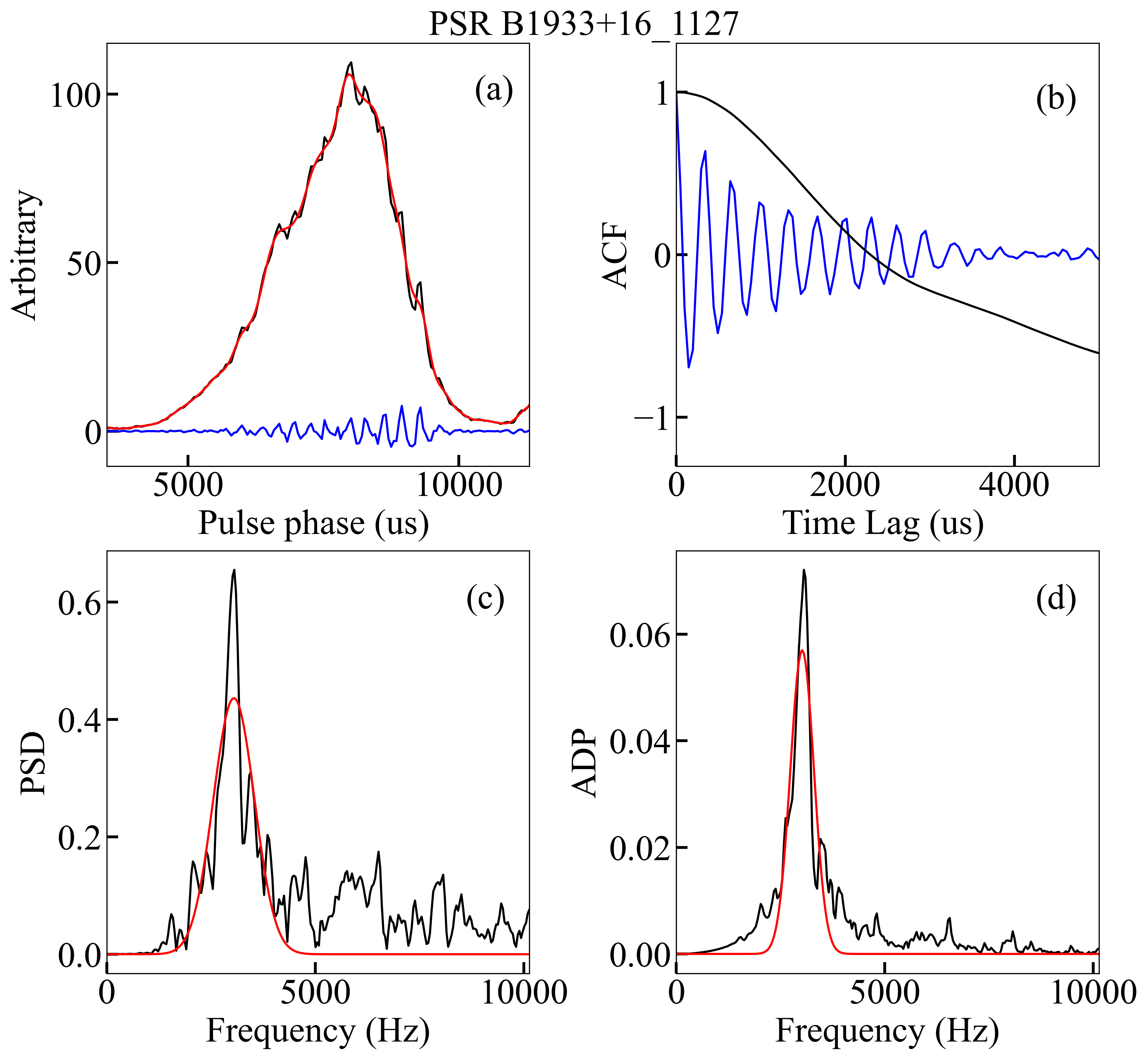}
    \caption{False positive sample 1127 } 
    \label{fig:pseudo_qmp2}
\end{subfigure}

\vspace{0.5cm}

\begin{subfigure}[b]{0.48\textwidth}
    \centering
    \includegraphics[width=\textwidth]{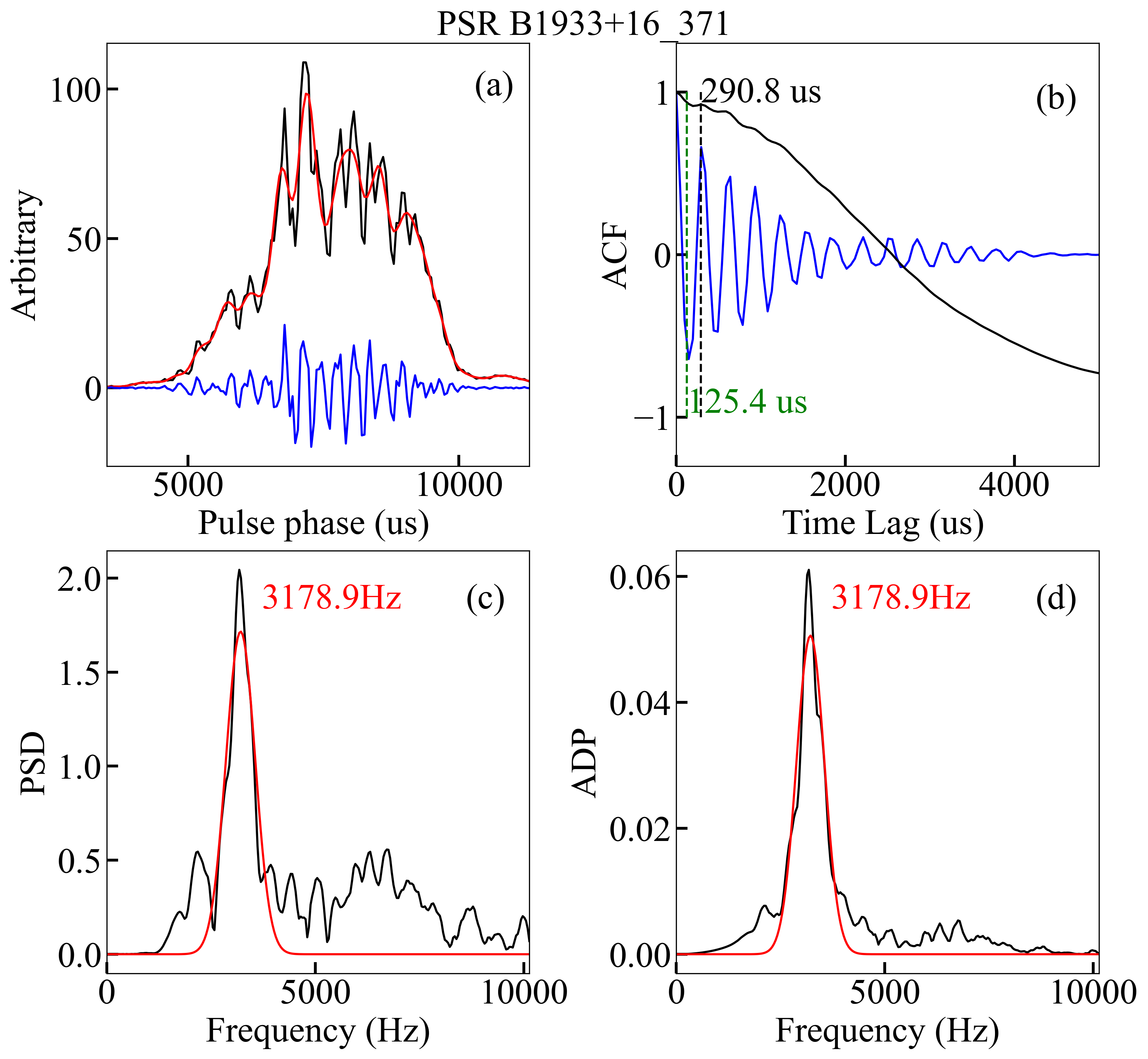}
    \caption{True positive sample 371 (SNR=34.18)} 
    \label{fig:positives1}
\end{subfigure}
\hfill
\begin{subfigure}[b]{0.48\textwidth}
    \centering
    \includegraphics[width=\textwidth]{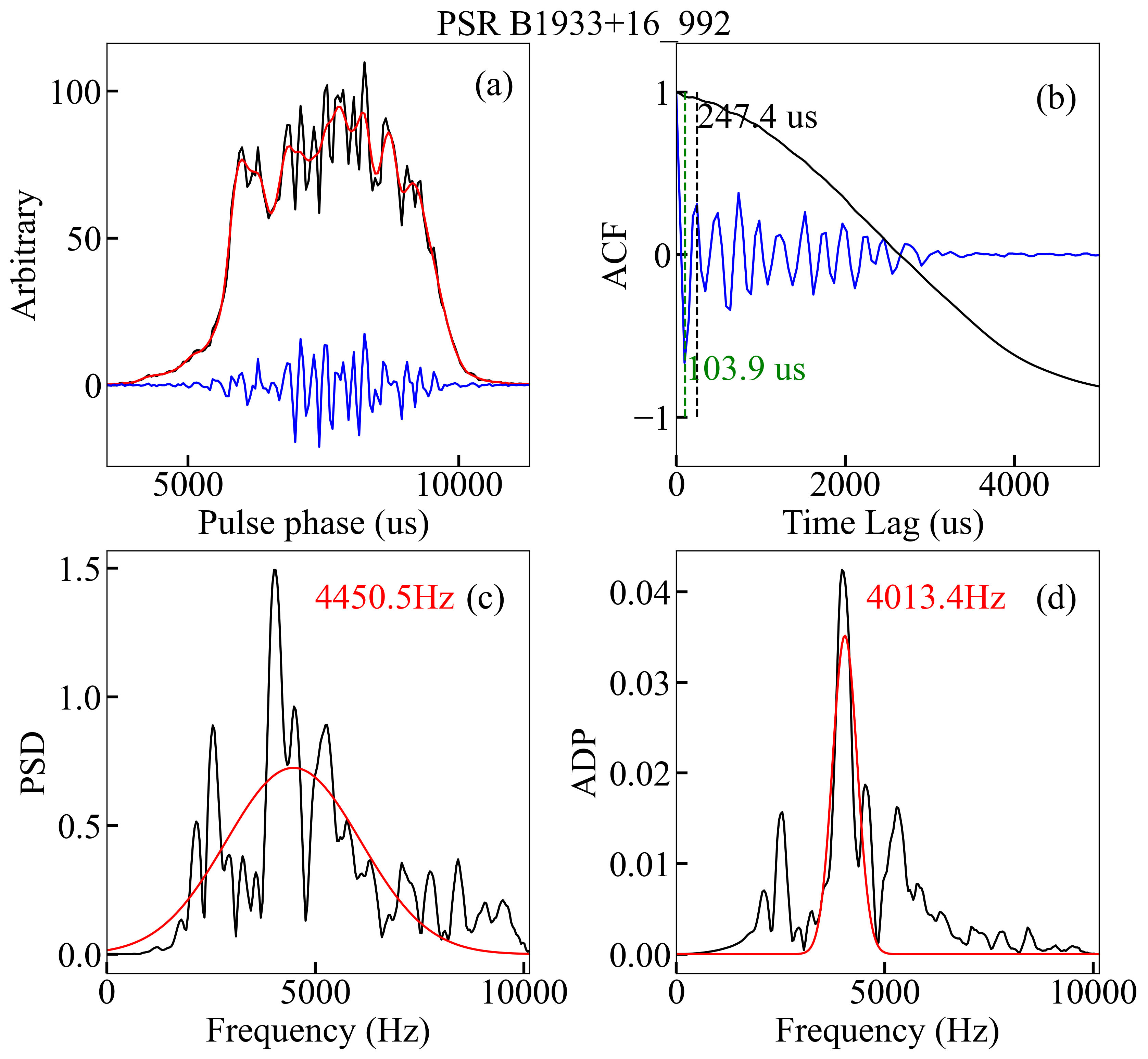}
    \caption{True positive sample 992 (SNR=10.16)} 
    \label{fig:positives2}
\end{subfigure}

\caption{
Diagnostic plot comparisons: 
\textbf{(i-ii)} Pseudo-quasi-periodic micropulses: Present in single-stage model identification;
\textbf{(iii-iv)} True positives - Correctly detected QMPs with characteristic unimodal ADP patterns and clear time-domain microstructures. SNR values denote signal-to-noise ratios.
}
\label{fig:true_positive_examples}
\end{figure*}

Despite both Model 1 (single-pulse image classification) and Model 2 (ADP image classification) demonstrating high precision exceeding 96.00\% on the validation set (red hollow circle in Figure.~\ref{fig:training_curves}), their performance on the PSR B1933+16 (2020) test set significantly degraded, with precisions of only 21.93\% and 52.36\%, respectively. This discrepancy stems from two critical misclassification types: First, certain single-pulse images exhibit visual structures in the time domain that resemble quasi-periodic micropulses (Figure.~\ref{fig:pseudo_qmp1}a), yet their corresponding ADP subplots (Figure.~\ref{fig:pseudo_qmp1}b-d) do not exhibit single-peaked periodic features, causing Model 1 to misclassify them as QMPs. 
Second, while some samples exhibit distinct periodic features in their ADP subplots (Figures~\ref{fig:pseudo_qmp2}b-d), their corresponding single-pulse images lack clustered microstructural signals. According to the criteria defined in Section~\ref{sec:classification}), such profiles should be classified as NQMPs. However, these non-clustered fluctuations in the single-pulse profiles—potentially representing low-amplitude stochastic variations—were misinterpreted by Model 2 as quasi-periodic microstructures based on their ADP characteristics, leading to false positive classifications. Therefore, the joint screening mechanism of the dual-stage model (Model 1 initial screening + Model 2 verification) is essential for significantly enhancing recognition reliability.

\subsection{Attribution of Missed and False Detections in the DSR Model}
\label{subsec:error_analysis}

Testing the DSR model on the PSR B1933+16 (2020) dataset revealed 30 missed detections and 33 false detections (confusion matrix in Figure~\ref{fig:confusion_matrix_example}). These errors primarily stem from image feature disparities and mild model overfitting. 

\begin{figure*}[ht!]
\centering
\captionsetup[subfigure]{labelformat=simple}
\renewcommand{\thesubfigure}{(\roman{subfigure})} 

\begin{subfigure}[b]{0.48\textwidth}
    \centering
    \includegraphics[width=\textwidth]{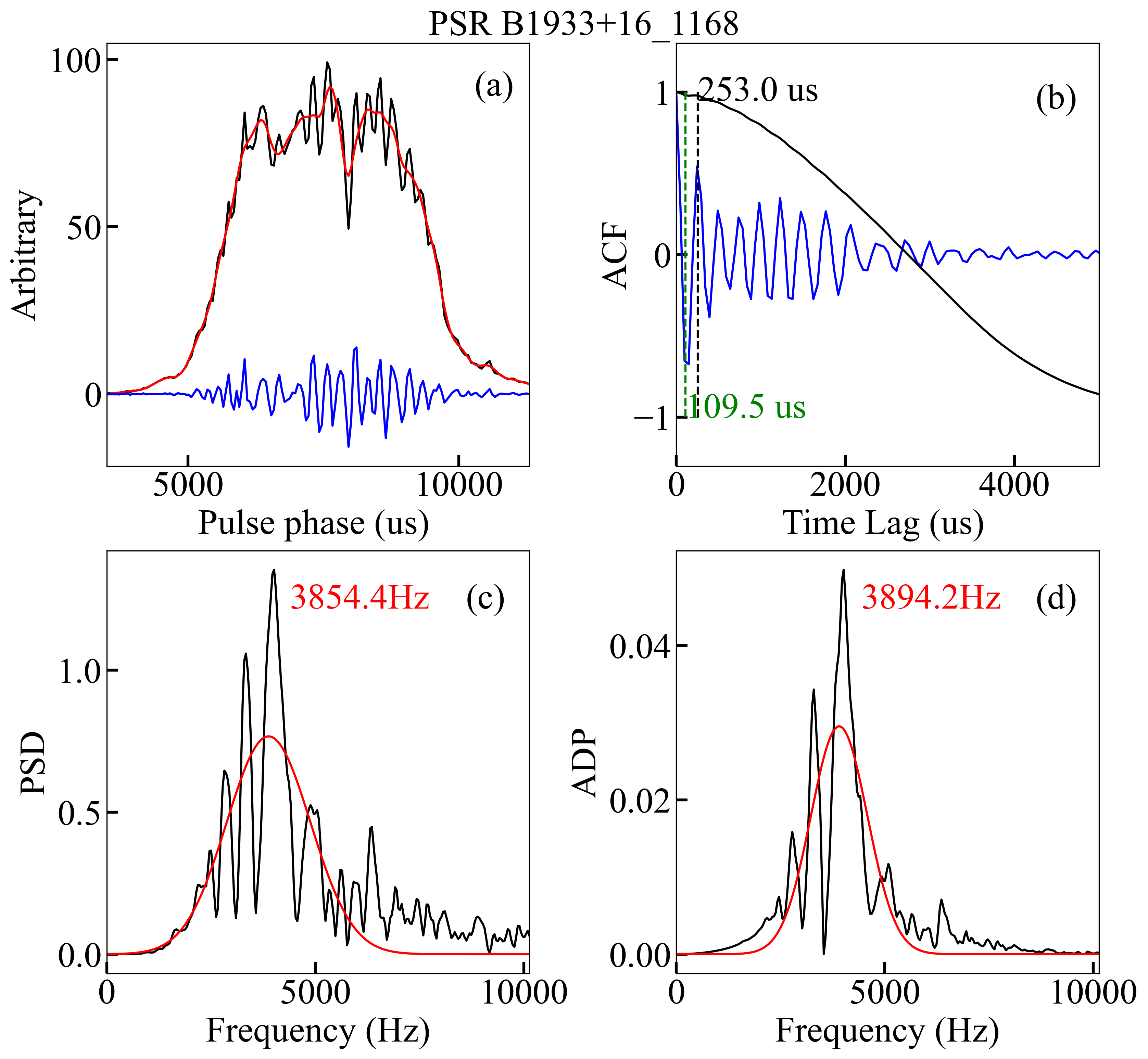}
    \caption{Missed QMP sample 1168 (SNR=38.82)} 
    \label{fig:missed_qmp1}
\end{subfigure}
\hfill
\begin{subfigure}[b]{0.48\textwidth}
    \centering
    \includegraphics[width=\textwidth]{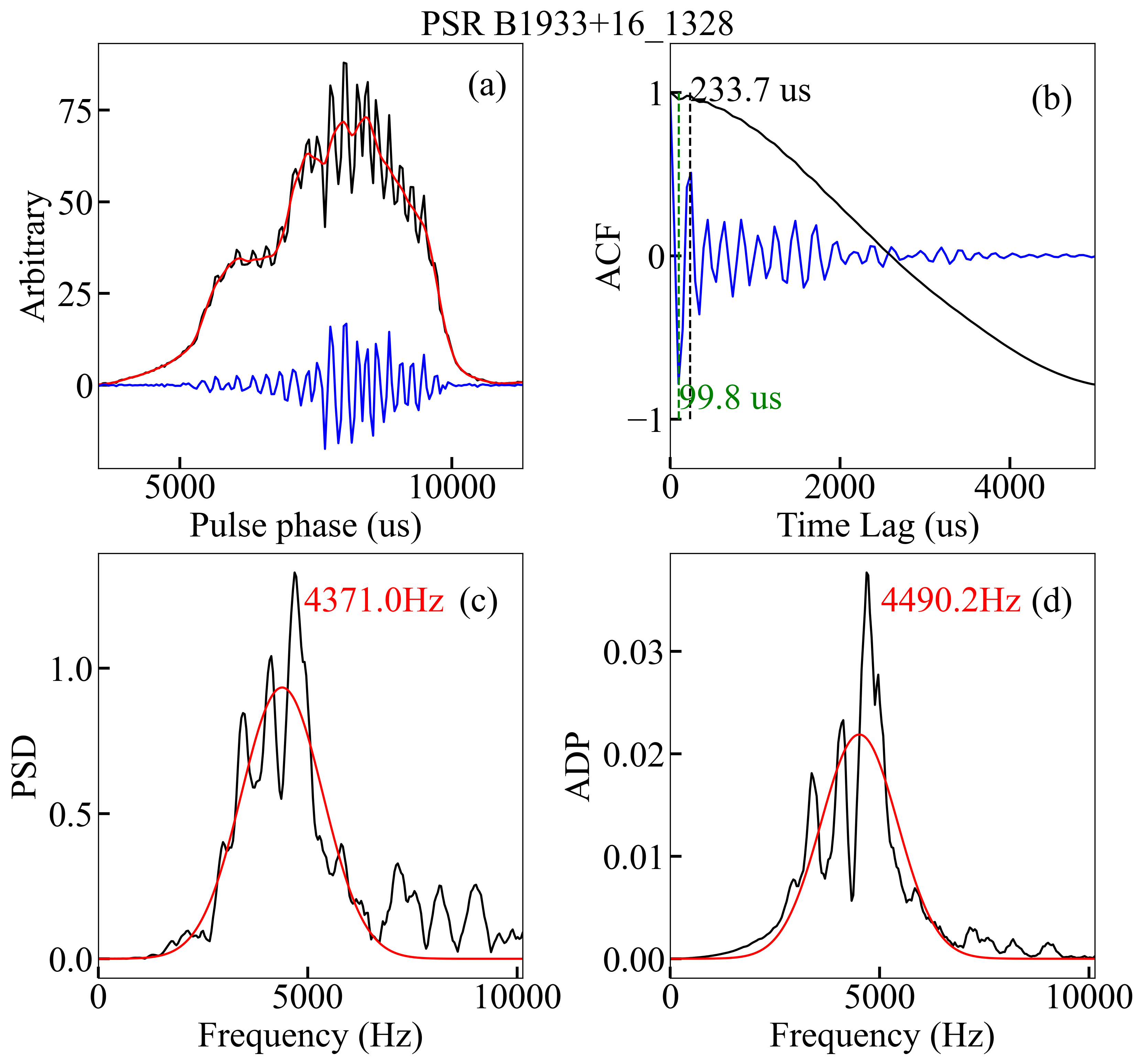}
    \caption{Missed QMP sample 1328 (SNR=49.48) } 
    \label{fig:missed_qmp2}
\end{subfigure}

\vspace{0.5cm}

\begin{subfigure}[b]{0.48\textwidth}
    \centering
    \includegraphics[width=\textwidth]{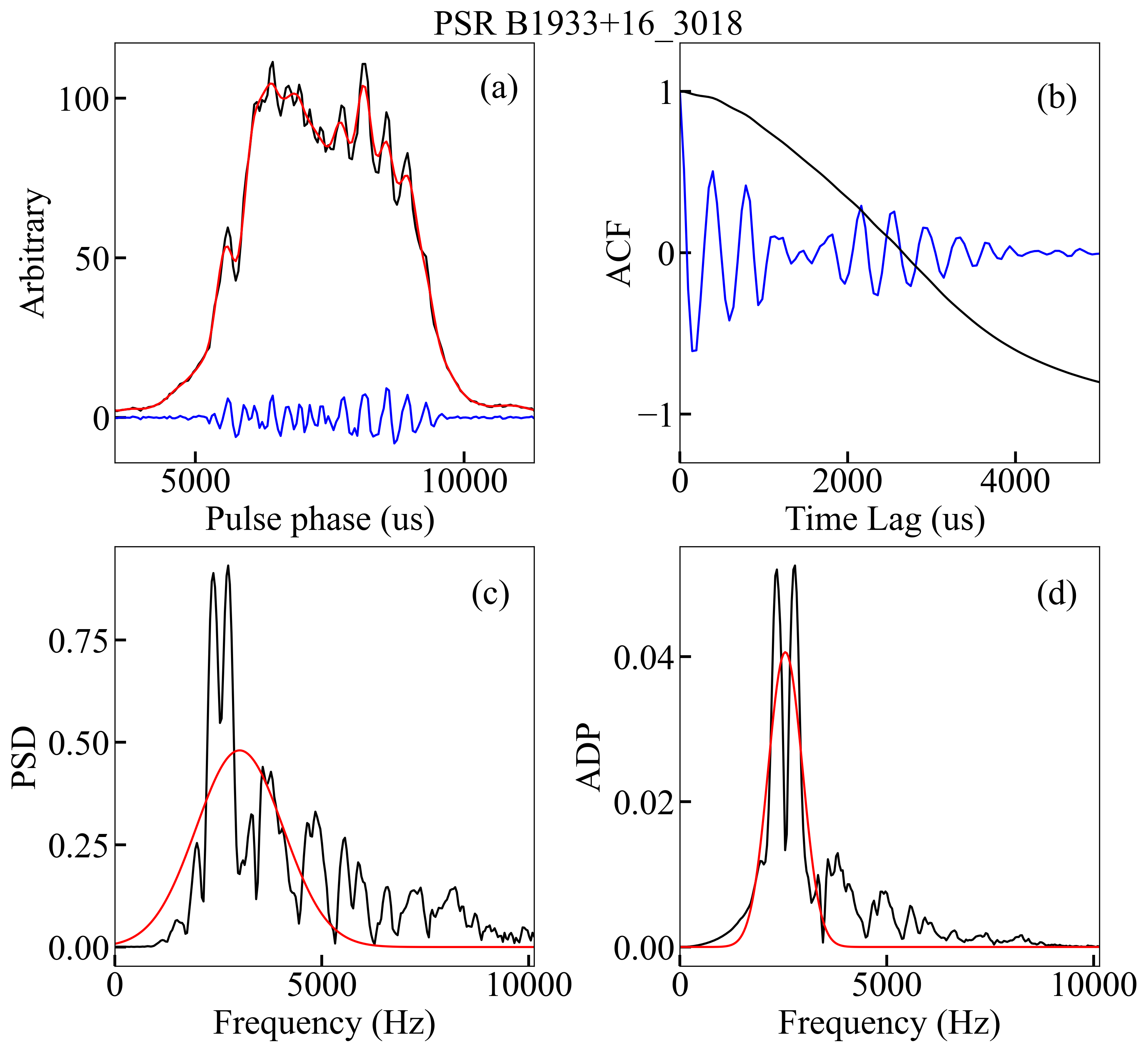}
    \caption{False detection sample 3018 (SNR=19.26) } 
    \label{fig:false_det1}
\end{subfigure}
\hfill
\begin{subfigure}[b]{0.48\textwidth}
    \centering
    \includegraphics[width=\textwidth]{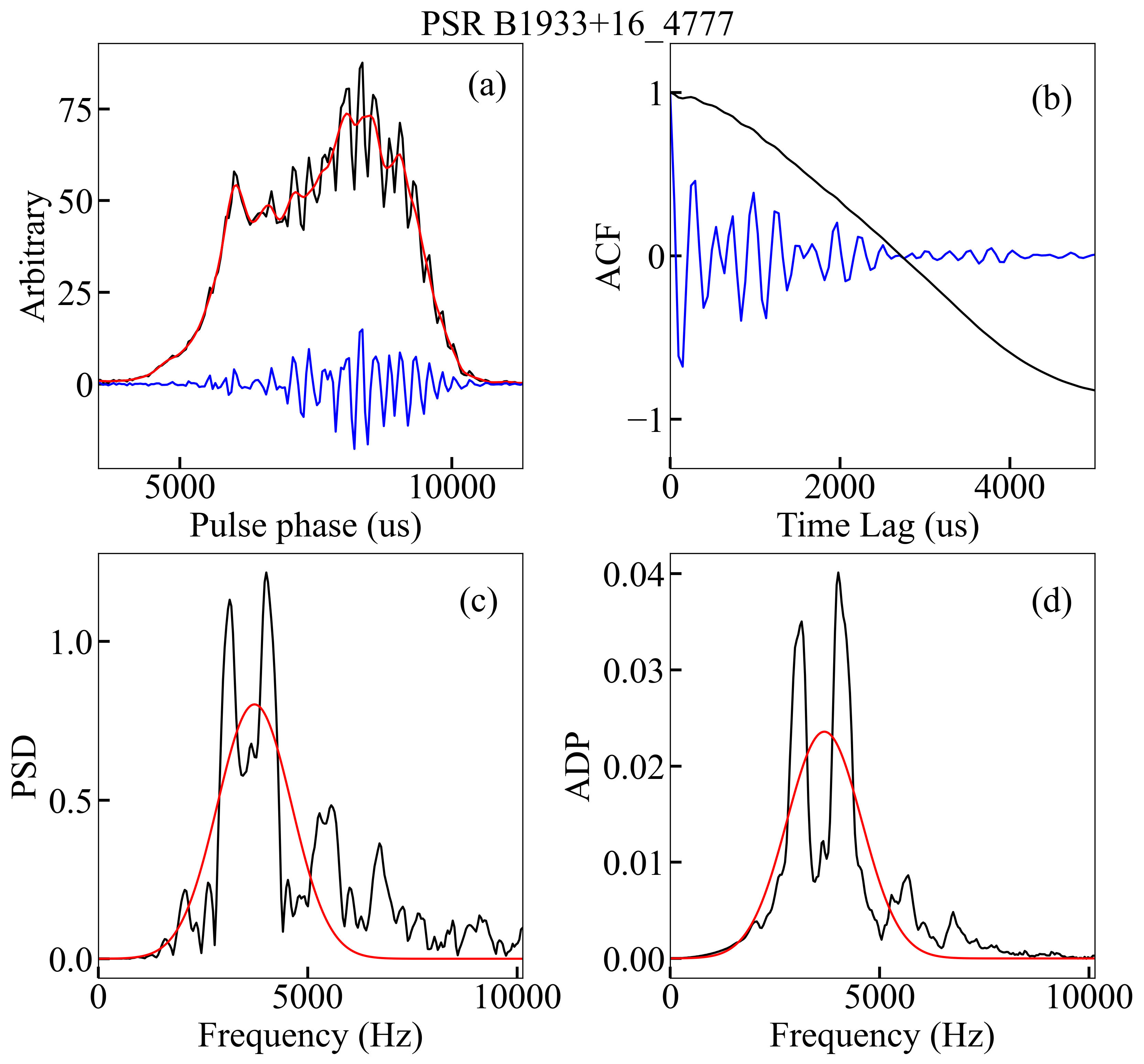}
    \caption{False detection sample 4777 (SNR=39.14) } 
    \label{fig:false_det2}
\end{subfigure}

\caption{
Error case analysis: 
\textbf{(i-ii)} Missed detections - QMPs with hybrid ADP patterns containing both unimodal peaks and dense clusters; 
\textbf{(iii-iv)} False detections: NQMPs misclassified as QMPs due to bimodal features; SNR values indicate signal-to-noise ratios.
}
\label{fig:error_examples}
\end{figure*}

For missed detections, samples exhibit significantly higher feature complexity than correctly identified positives (e.g., Figure~\ref{fig:positives1}-\ref{fig:positives2}): although their single-pulse images (panels (a) in Figure~\ref{fig:missed_qmp1} and Figure~\ref{fig:missed_qmp2}) show clear temporal microstructures and are accurately classified by Model 1, the corresponding high-SNR ADP images (SNR \textgreater 10, panels (d) in the same figures) display hybrid modality signatures—simultaneously containing characteristic high-amplitude unimodal peaks of QMPs and typical low-amplitude dense peak clusters of NQMPs. This feature ambiguity causes Model 2 to misclassify them as NQMPs due to insufficient classification confidence. For false detections, while single-pulse images are correctly identified by Model 1, their high-SNR ADP images (SNR \textgreater 10) present clean backgrounds resembling unimodal patterns but contain distinct bimodal structures (likely corresponding to two independent periodic systems rather than target quasi-periodic features), as shown in Figure~\ref{fig:false_det1} and Figure~\ref{fig:false_det2}. These borderline cases lead Model 2 to misclassify them as QMPs.

Concurrently, 
the training loss curves of the DSR (Figure~\ref{fig:training_curves}) exhibit characteristics of mild overfitting. Specifically, although the model achieves strong classification accuracy on both the training and validation sets, a noticeable gap persists between their respective loss values, indicating a potential tendency toward overfitting. Despite the implementation of an early stopping strategy and the convergence of accuracy, a persistent loss gap remains at the respective optimal performance points of the two models by epoch 100 (0.2462 for Model 1 and 0.2221 for Model 2), which somewhat constrains the model's ability to represent complex features. Furthermore, limited training dataset scale—particularly scarce composite ADP patterns and multi-periodic structure samples—further diminishes generalization capacity for atypical features, constituting the fundamental intrinsic error source. These borderline cases collectively demonstrate that enhancing the model's capacity to characterize composite ADP patterns will be crucial for improving generalization performance in future iterations.

\subsection{Distribution Characteristics of the Quasi-periodic Micropulse Period $P_\mu$}
\label{subsec:p_mu_distribution}

\begin{figure*}[htbp]
\centering
\captionsetup[subfigure]{labelformat=simple, labelsep=colon}
\renewcommand{\thesubfigure}{(\alph{subfigure})}

\begin{subfigure}[b]{0.40\textwidth}
    \centering
    \includegraphics[width=\linewidth]{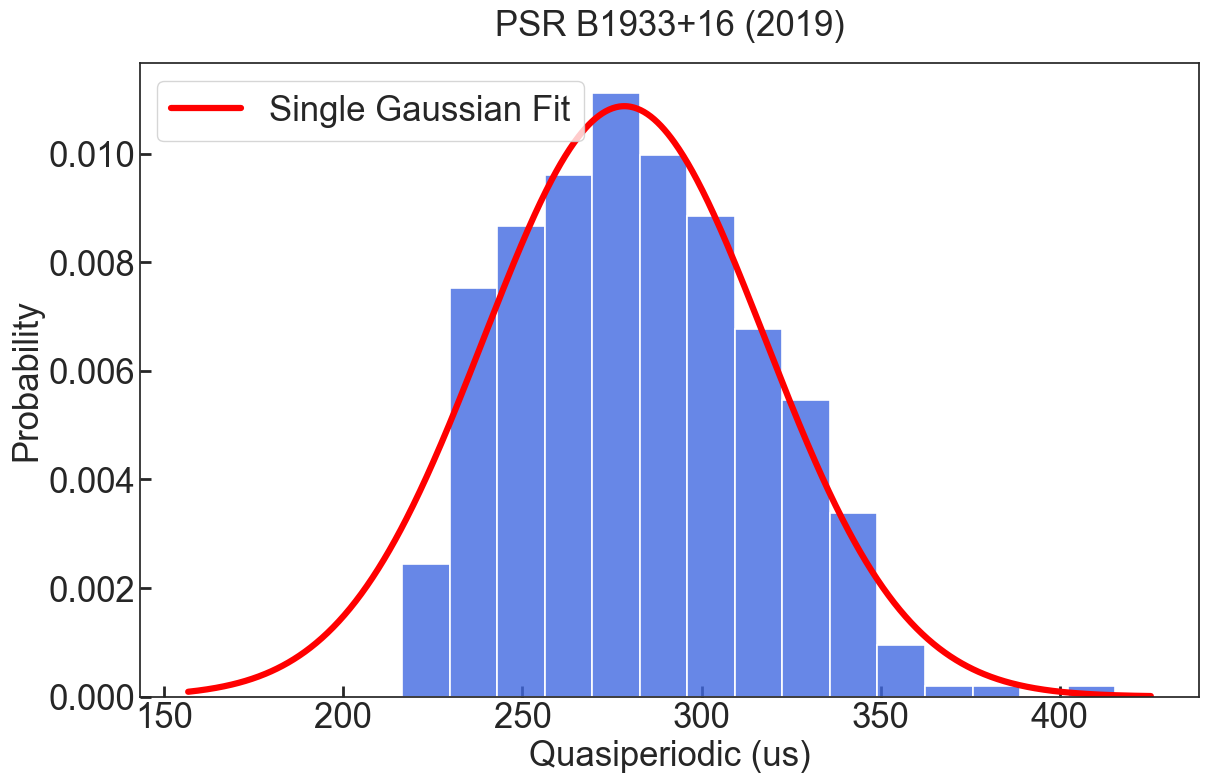}
    \caption{}
    \label{fig:p_mu_2019}
\end{subfigure}
\hfill
\begin{subfigure}[b]{0.40\textwidth}
    \centering
    \includegraphics[width=\linewidth]{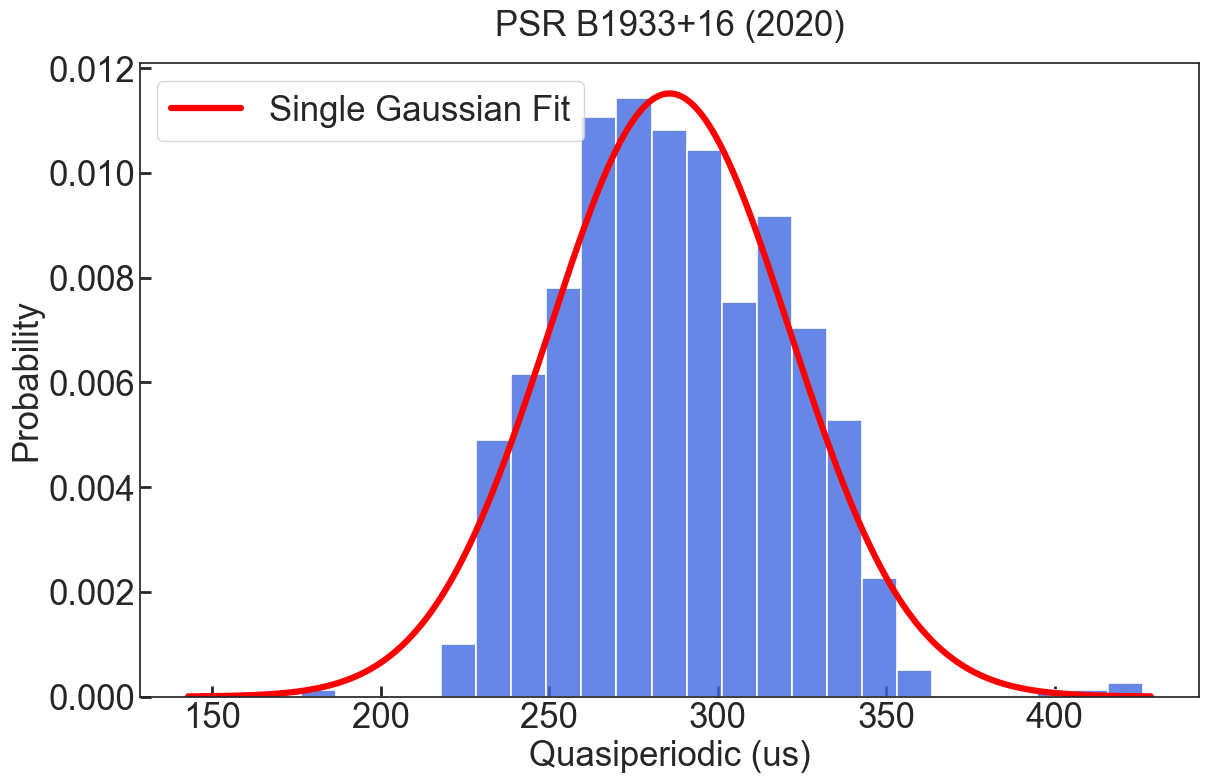}
    \caption{}
    \label{fig:p_mu_2020}
\end{subfigure}

\vspace{0.3cm} 

\begin{subfigure}[b]{0.40\textwidth}
    \centering
    \includegraphics[width=\linewidth]{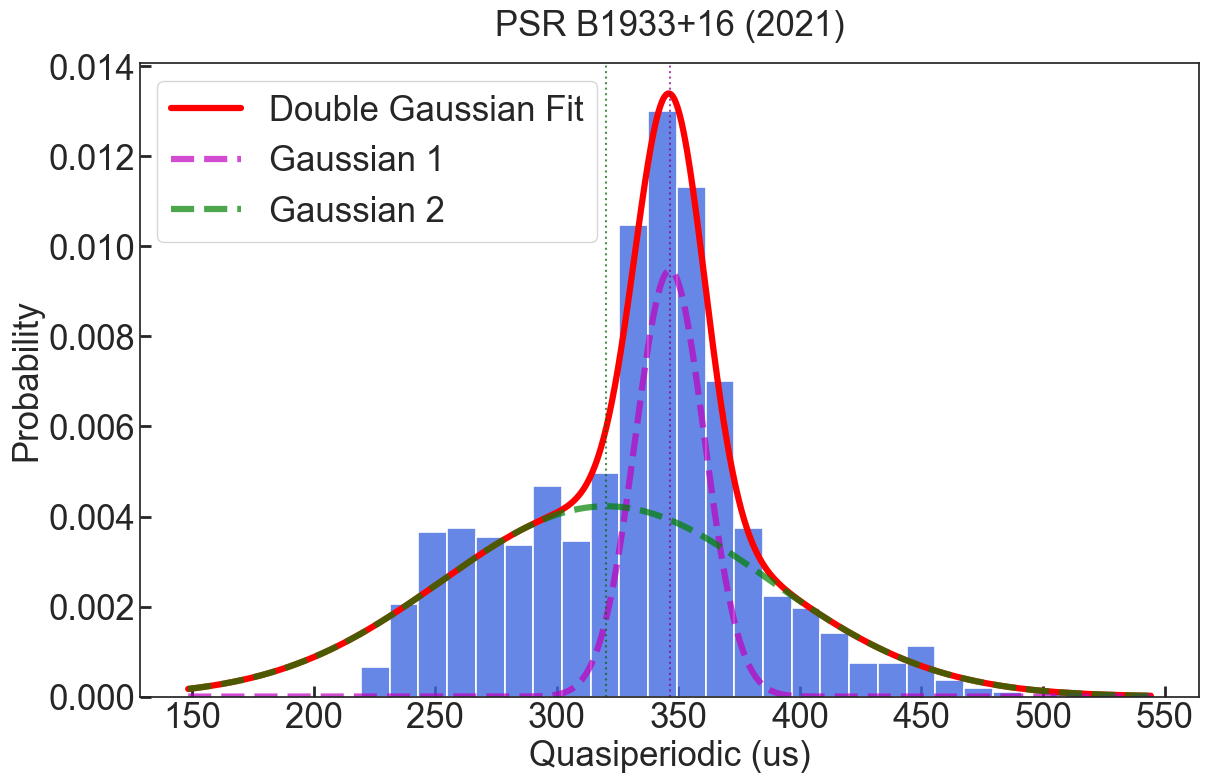}
    \caption{}
    \label{fig:p_mu_2021}
\end{subfigure}
\hfill
\begin{subfigure}[b]{0.40\textwidth}
    \centering
    \includegraphics[width=\linewidth]{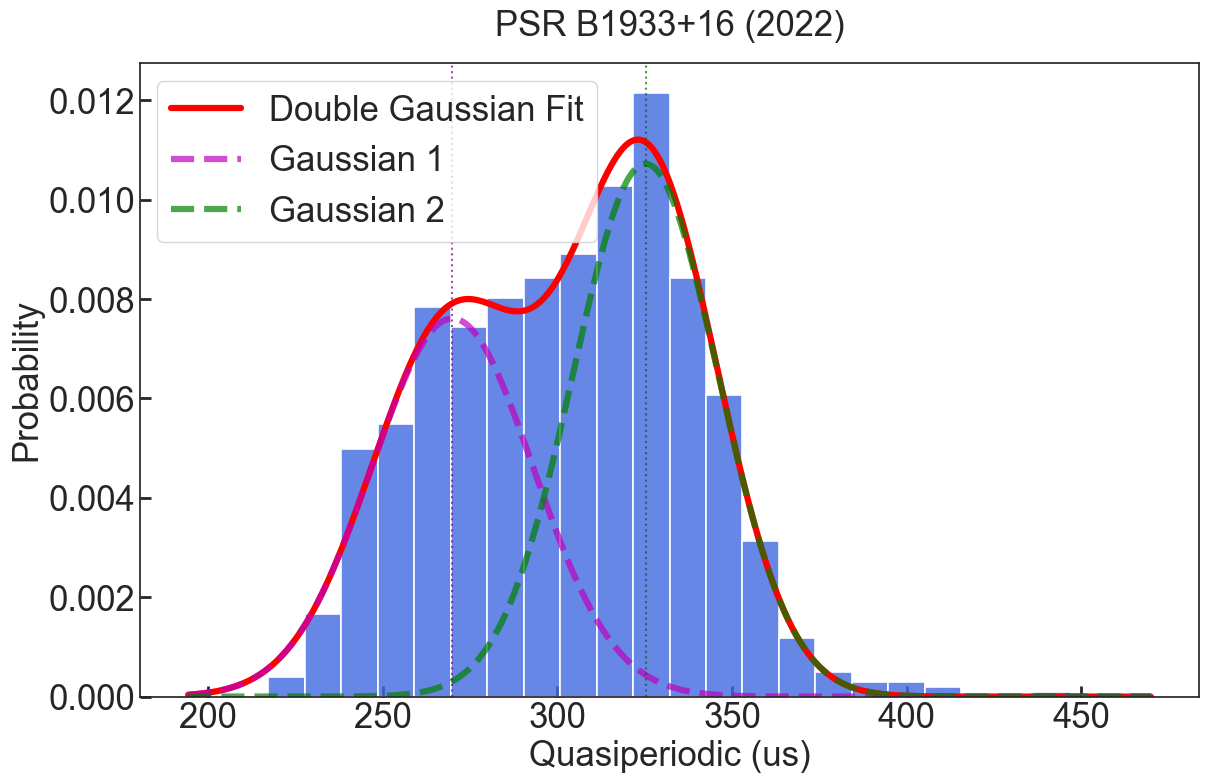}
    \caption{}
    \label{fig:p_mu_2022}
\end{subfigure}

\vspace{0.3cm} 

\begin{subfigure}[b]{0.40\textwidth}
    \centering
    \includegraphics[width=\linewidth]{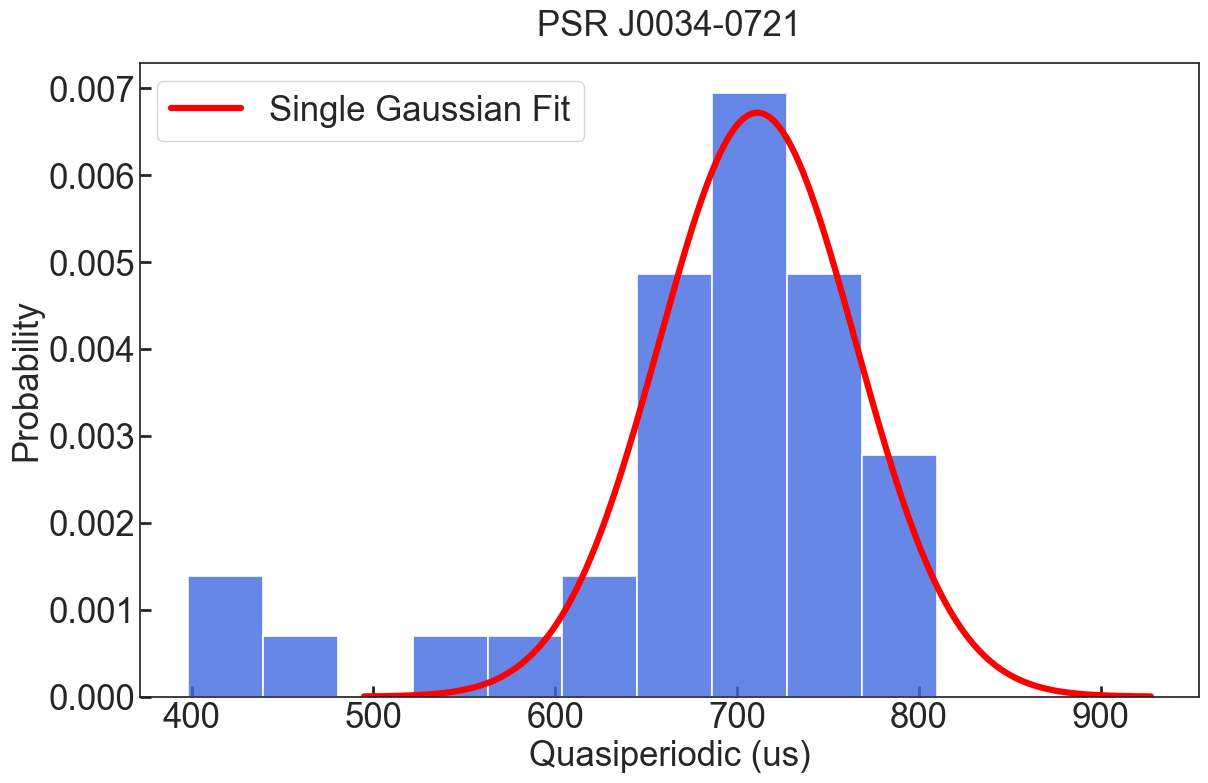}
    \caption{}
    \label{fig:p_mu_j0034}
\end{subfigure}
\hfill
\begin{subfigure}[b]{0.40\textwidth}
    \centering
    \includegraphics[width=\linewidth]{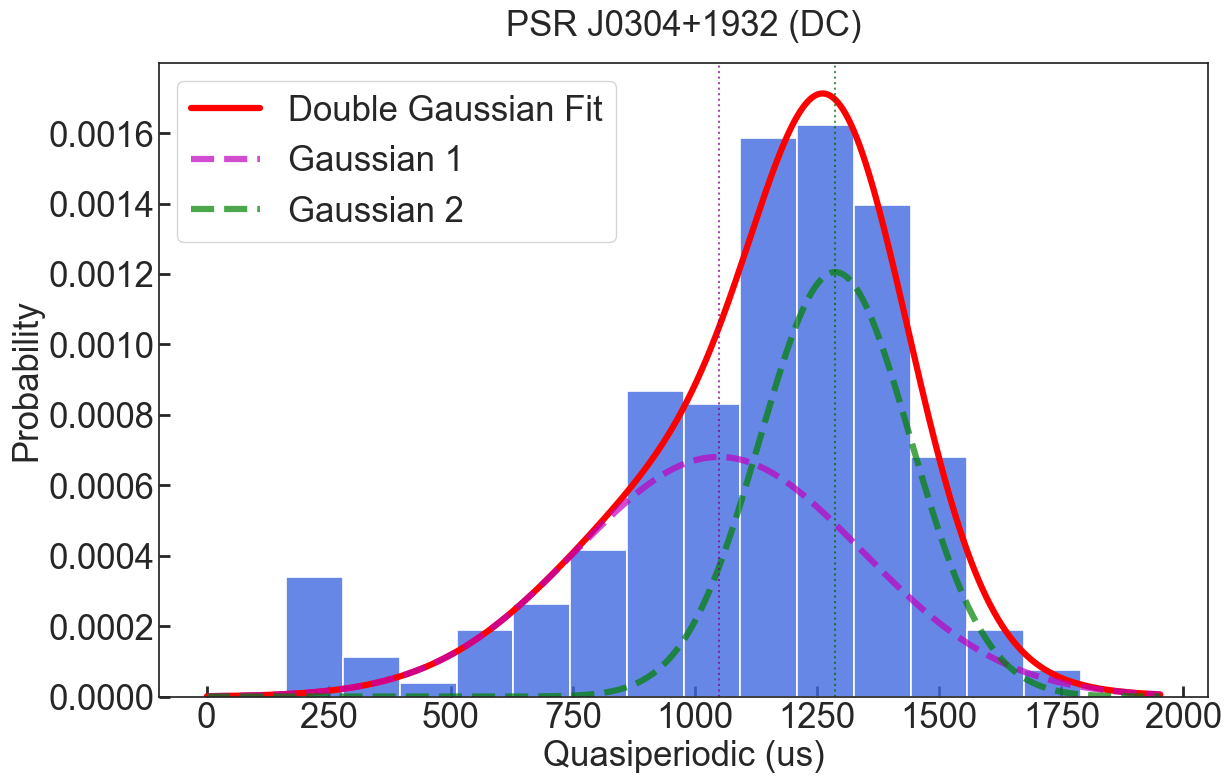}
    \caption{}
    \label{fig:p_mu_j0304_dc}
\end{subfigure}

\vspace{0.3cm} 

\centering
\begin{subfigure}[b]{0.40\textwidth} 
    \centering
    \includegraphics[width=\linewidth]{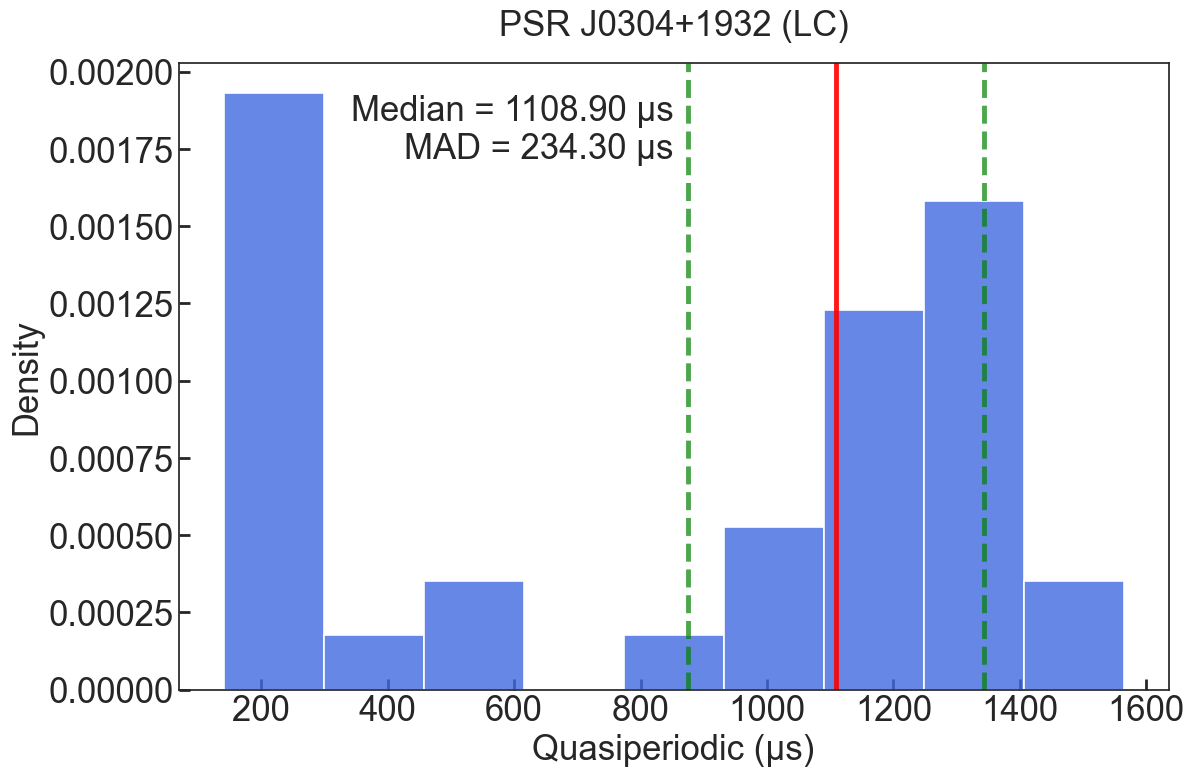}
    \caption{}
    \label{fig:p_mu_j0304_lc}
\end{subfigure}

\caption{%
Statistical distribution of $P_\mu$ parameters for multiple pulsars: 
(a) Known positive samples from PSR B1933+16 (2019), (b) True positive samples identified by DSR in PSR B1933+16 (2020),
(c) True positive samples identified by DSR in PSR B1933+16 (2021),
(d) True positive samples identified by DSR in PSR B1933+16 (2022), 
(e) Samples from PSR J0034-0721,
(f) Samples from the DC of PSR J0304+1932,
(g) Samples from the LC of PSR J0304+1932.
The horizontal axis represents $P_\mu$ ($\mu$s), and the vertical axis shows the sample count. 
Panels (a) to (f) show Gaussian fitting results, while panel (g) exhibits a more discrete distribution.
}
\label{fig:p_mu_distribution}
\end{figure*}

Based on the test results from the DSR model (Section~\ref{sec:test}), we extracted true positive QMP samples of PSR B1933+16 identified during 2020-2022, with sample sizes of 764, 906, and 979 respectively. Following the procedure illustrated in Figure~\ref{fig:diagnostic}, diagnostic plots were generated for annual QMP detections, their ACF was computed, and the period corresponding to the first ACF peak was recorded as $P_\mu$ (exemplified by 240.1 $\mu$s in Figure~\ref{fig:diagnostic}(b)). By comparing the $P_\mu$ distributions of the known 400 QMP samples from 2019 with those identified by DSR during 2020-2022 (Figure~\ref{fig:p_mu_distribution}), the Freedman-Diaconis rule \citep{freedman1981histogram} was employed to determine histogram bin counts (Table~\ref{tab:performance_Gaussian_Fitting}) with Gaussian fitting applied. 

The results reveal that the 2020 data follows a single Gaussian distribution ($285.66 \pm 42.05\,\mu\text{s}$), the 2021 data exhibits a bimodal distribution with peaks at $320.31 \pm 79.85\,\mu\text{s}$ and $346.51 \pm 16.76\,\mu\text{s}$, and the 2022 data similarly shows bimodality with characteristic values of $269.90 \pm 27.26\,\mu\text{s}$ and $325.23 \pm 24.45\,\mu\text{s}$. All true positive micropulse $P_\mu$ values are concentrated within the full width at half maximum (FWHM) of their corresponding Gaussian peaks and demonstrate consistency within errors with the characteristic quasi-periodic micropulse range ($400 \pm 200\,\mu\text{s}$) reported for PSR B1933+16 by \citet{Mitra2016}. The observed year-to-year fluctuations in the $P_\mu$ distribution may be attributed to strong diffractive interstellar scintillation. As demonstrated by \citet{Wang2018A&A...618A.186W} for this same pulsar, scintillation parameters ($\Delta\nu_d$, $\Delta t_d$) vary significantly between epochs, indicating that the line of sight traverses different turbulent interstellar structures over time, which can modulate the observed micropulse structure from year to year.

Further analysis of cross-pulsar samples indicates: The $P_\mu$ of 35 true positive QMP samples from PSR J0034-0721 follows a single Gaussian distribution ($711.22 \pm 63.70\,\mu\text{s}$), consistent within errors with the range (500-5500 $\mu$s) reported by \citet{Wen2021ApJ...918...57W}. The high sensitivity of FAST provided high-SNR data, which not only reduced individual measurement errors but also enabled the reliable identification of QMPs from a significantly larger initial sample (3118 pulses in this work vs. 146 in \citet{Wen2021ApJ...918...57W}), leading to our more precise measurement of $P_{\mu}$. simultaneously, 228 samples from the DC of PSR J0304+1932 conform to a bimodal Gaussian distribution ($1048.99 \pm 345.71\,\mu\text{s}$ and $1287.18 \pm 181.95\,\mu\text{s}$), while the 36 samples from its LC exhibit a discrete distribution. For the LC samples, we computed the median and median absolute deviation (MAD) of the distribution as shown in the figure to determine $P_\mu$ ($1108.90 \pm 234.30\,\mu\text{s}$). The dominant component shows consistency within errors with the DC characteristic range ($1490 \pm 570\,\mu\text{s}$) reported by \citet{Mitra2015ApJ...806..236M}, and the leading component with the LC range ($1730 \pm 970\,\mu\text{s}$). In summary, statistical analysis of $P_\mu$ from cross-year and cross-pulsar true positive QMP samples confirms the effectiveness of DSR in identifying quasi-periodic microstructural features, while minimizing human intervention to significantly enhance analytical efficiency.

\begin{deluxetable*}{ccccc}[!ht]
\digitalasset
\tablewidth{0pt}
\tablecaption{Distribution of Quasi-periodic Values for QMP Samples
\label{tab:performance_Gaussian_Fitting}}
\tablehead{
\colhead{Source} & \colhead{Sample Size} & \colhead{F-D bins} & \colhead{Quasi-periodic} & \colhead{FWHM} \\
\colhead{}&\colhead{} & \colhead{}& \colhead{($\mu$s)} & \colhead{($\mu$s)} \\
}
\startdata
PSR B1933+16 (2019) & 400 & 15 & \(278.45 \pm 46.24\) & 92.4835 \\
PSR B1933+16 (2020) & 764 & 24 & \(285.66 \pm 42.05\) & 84.1075 \\
{} -- {} & -- & -- & \(346.51 \pm 16.76\) & 33.5295 \\
\addlinespace
PSR B1933+16 (2021) & 906 & 25 & \(320.31 \pm 79.85\) & 159.6970 \\
{} -- {} & -- & -- & \(346.51 \pm 16.76\) & 33.5295 \\
\addlinespace
PSR B1933+16 (2022) & 979 & 22 & \(269.90 \pm 27.26\) & 54.5120 \\
{} -- {} & -- & -- & \(325.23 \pm 24.45\) & 48.8950 \\
\addlinespace
PSR J0034-0721 & 35 & 10 & \(711.22 \pm 63.70\) & 127.4070 \\
\addlinespace
PSR J0304+1932 (DC) & 228 & 14 & \(1048.99 \pm 345.71\) & 691.4112 \\
{} -- {} & -- & -- & \(1287.18 \pm 181.95\) & 363.8902 \\
\addlinespace
PSR J0304+1932 (LC) & 36 & 9 & \(1108.90 \pm 234.30\) & -- \\
\enddata
\tablecomments{Most quasi-periodic values are expressed as $\mu \pm \frac{\text{FWHM}}{2}$ (mean $\pm$ half of the full width at half maximum), while the quasi-periodic value for PSR J0304+1932 (LC) is given as median ± MAD.}
\end{deluxetable*}

\subsection{Astrophysical Implications of Bimodal Quasi-periodic Distributions}

In this study, we detected QMP in PSR B1933+16 (2021--2022 observation years) and in the DC of PSR J0304+1932, revealing a bimodal distribution structure in their quasi-periodicity ($P_\mu$) (Figure~\ref{fig:p_mu_distribution}(c,d,f))). This finding has significant implications for our understanding of the microphysical nature of the source's emission. It is important to emphasize that bimodal distributions of QMP periods are not an isolated phenomenon, but rather a common feature observed across various types of pulsars. In their unifying study of all types of radio-emitting neutron stars (from millisecond pulsars to magnetars), \citet{Kramer2024NatAs...8..230K} clearly demonstrated that quasi-periodic fluctuations are ubiquitous, and specifically reported a clear "bimodal distribution" in the magnetar XTE J1810-197. Furthermore, in millisecond pulsars, while \citet{Liu2022MNRAS.513.4037L} did not explicitly report bimodality for PSR J1022+1001, PSR J2145-0750, and PSR J1744-1134, their period distributions (their Figure 3) exhibit broad ranges and complex morphologies beyond single-peaked structures, suggesting the potential presence of multiple characteristic emission scales in their magnetospheres as well. This quasi-periodic variability and multi-peaked behavior across different pulsar types and rotation periods strongly suggests a common physical origin behind these phenomena.

\citet{Kramer2024NatAs...8..230K} established a core interpretive framework for this phenomenon: quasi-periodic substructures result from a series of discrete ``beamlets" (angular patterns) with characteristic angular widths sweeping across the observer's line of sight—a geometric effect. To explain the commonly observed quasi-periodic fluctuations and multi-peaked distributions, they proposed an unified physical explanation: the current distribution in pulsar magnetospheres universally contains structures of different scales. This picture is supported by the theoretical model of \citet{Thompson2022_1,Thompson2022_2ApJ...933..232T}, which links radio emission to slow tearing mode instabilities in the open field line region of the magnetosphere and naturally predicts the formation of multiple characteristic scales in the current distribution.

Within this universal theoretical framework, the bimodal distributions we observed in PSR B1933+16 and the DC of PSR J0304+1932 provide direct observational evidence for the simultaneous presence of multiple dominant characteristic scales in their magnetospheres. These different scales may correspond to various modes of instability within the current sheets, or to multiple dominant emission regions of specific sizes within the magnetosphere.


\section{Conclusion} \label{sec:con}
\label{sec:conclusion}

This study designed and validated a DSR framework for automated detection of QMPs in pulsars. Firstly, the DSR model trained on 2019 observational data from PSR B1933+16 achieved 95.85\% precision, 96.10\% recall, and 95.97\% F1-score on the manually labeled 2020 test set; attained precisions of 95.16\% and 96.17\% respectively on unlabeled 2021--2022 test sets; achieved 94.59\% precision on the PSR J0034-0721 test set; and obtained 93.06\% and 94.73\% precision on the DC and LC of the two-component pulsar PSR J0304+1932 (average precision of 94.87\% across 5 unlabeled test sets). Secondly, the periodicity period $P_\mu$ of genuine QMPs detected by DSR demonstrates agreement within errors with prior studies:
$P_\mu$ values for PSR B1933+16 (2020--2022) ($285.66 \pm 42.05\,\mu\text{s}$, $320.31 \pm 79.85\,\mu\text{s}$, $346.51 \pm 16.76\,\mu\text{s}$, $269.90 \pm 27.26\,\mu\text{s}$, and $325.23 \pm 24.45\,\mu\text{s}$) agree within errors with the characteristic range ($400 \pm 200\,\mu\text{s}$) reported by \citet{Mitra2016}; $P_\mu = 711.22 \pm 63.70\,\mu\text{s}$ for PSR J0034-0721 is consistent with \citet{Wen2021ApJ...918...57W} ($500$--$5500\,\mu\text{s}$); while $P_\mu$ values for the DC of PSR J0304+1932 ($1048.99 \pm 345.71\,\mu\text{s}$, $1287.18 \pm 181.95\,\mu\text{s}$) and the LC ($1108.90 \pm 234.30\,\mu\text{s}$) respectively match \citet{Mitra2015ApJ...806..236M} reported values ($1490 \pm 570\,\mu\text{s}$) and ($1730 \pm 970\,\mu\text{s}$).

Experimental results confirm that DSR exhibits exceptional QMP recognition capability across multi-year datasets, cross-source pulsars (PSR B1933+16, PSR J0034-0721, PSR J0304+1932), and multi-component pulsars (PSR J0304+1932 DC/LC), significantly outperforming manual screening efficiency and demonstrating the feasibility of this approach. 
The primary limitation lies in the training data being predominantly from a single pulsar (PSR B1933+16), and the DSR architecture's limited capability in handling borderline cases. Future work will focus on expanding the training dataset and exploring end-to-end models to further optimize performance, ultimately aiming to construct an automated pipeline for massive data from facilities like FAST.

\section*{Acknowledgments}
We hereby acknowledge support from the following agencies and programs: National Natural Science Foundation of China (Grant Nos. 12563008, 11988101, U1731238, U2031117, 11565010, 11725313, 1227308, 12041303, 12588202); National Key R\&D Program of China (No. 2023YFE0110500); National SKA Program of China (Nos. 2020SKA0120200, 2022SKA0130100, 2022SKA0130104); the Foundation of Guizhou Provincial Education Department (No. KY(2023)059); Youth Innovation Promotion Association CAS (ID: 2021055); Youth Scientists Project of Basic Research in CAS (YSBR-006); Foreign Talents Program (No. QN2023061004L; E.G.); CAS Youth Interdisciplinary Team; Liupanshui Science and Technology Development Project (No. 52020-2024-PT-01);Foundation Research Project of Guizhou Provincial Association of Science and Technology (No. 2025XZQYXM-02-10); and the Cultivation Project for FAST Scientific Payoff and Research Achievement of CAMS-CAS. Yong-Kun Zhang is supported by the Postdoctoral Fellowship Program and China Postdoctoral Science Foundation under Grant Number BX20250158. P.W. acknowledges additional support from the CAS Youth Interdisciplinary Team, Youth Innovation Promotion Association CAS, and the Cultivation Project for FAST Scientific Payoff and Research Achievement of CAMS-CAS. D.L. is supported as a New Cornerstone Investigator. These supports were instrumental in the successful completion of this work.







\bibliography{Reference}{}
\bibliographystyle{aasjournalv7}



\end{document}